\documentclass[11pt,onecolumn]{IEEEtran} 

\ifCLASSOPTIONcompsoc
\else
\fi
\ifCLASSINFOpdf
   \usepackage[pdftex]{graphicx}
\else
   \usepackage[dvips]{graphicx}
\fi
\usepackage{amsmath,amssymb,amsthm,amsfonts,dsfont,stmaryrd,amsopn}
\usepackage{multirow}
\usepackage{colortbl}
\usepackage{hyperref}
\usepackage{rotating}
\usepackage{algorithm,algorithmic}

\newtheorem{theorem}{Theorem}[section]

\newtheorem{lemma}[theorem]{Lemma}

\ifCLASSOPTIONcompsoc
  \usepackage[caption=false,font=normalsize,labelfont=sf,textfont=sf]{subfig}
\else
  \usepackage[caption=false,font=footnotesize]{subfig}
\def\x{{\mathbf x}}

\def\w{{\mathbf w}}

\def\X{\mathbf{X}}
\def\Ph{\mathbf{\Phi}}
\def\Z{\mathbf{Z}}
\def\A{\mathbf{A}}
\def\B{\mathbf{B}}
\def\G{\mathbf{G}}
\def\W{\mathbf{W}}
\def\Y{\mathbf{Y}}
\def\U{\mathbf{U}}
\def\V{\mathbf{V}}
\def\H{\mathbf{H}}
\def\E{\mathbf{E}}
\def\I{\mathbf{I}}
\def\F{\mathbf{F}}
\def\D{\mathbf{D}}
\def\S{\mathbf{S}}
\def\0{\mathbf{0}}
\def\P{\mathbf{P}}
\def\Q{\mathbf{Q}}
\def\R{\mathbf{R}}
\def\Gam{\mathbf{\Gamma}}

\def\tX{\underline{\mathbf{X}}}
\def\tZ{\underline{\mathbf{Z}}}
\def\tG{\underline{\mathbf{G}}}
\def\tE{\underline{\mathbf{E}}}
\def\tW{\underline{\mathbf{W}}}
\def\tY{\underline{\mathbf{Y}}}
\def\tB{\underline{\mathbf{B}}}
\def\tH{\underline{\mathbf{H}}}
\def\t0{\underline{\mathbf{0}}}
\def\tP{\underline{\mathbf{P}}}
\def\tQ{\underline{\mathbf{Q}}}
\def\tR{\underline{\mathbf{R}}}

\def\tGamma{\underline{\mathbf{\Gamma}}}
\def\tOmega{\underline{\mathbf{\Omega}}}

\def\R{\mathds{R}} 

\DeclareMathOperator{\rank}{rank} 

\begin{document}
%
\title{Stable, Robust and Super Fast Reconstruction of Tensors Using Multi-Way Projections}
%
%
%
%

\author{Cesar~F.~Caiafa$^*$,~\IEEEmembership{Member,~IEEE,}
        and~Andrzej~Cichocki,~\IEEEmembership{Fellow,~IEEE}
\IEEEcompsocitemizethanks{\IEEEcompsocthanksitem Cesar~F.~Caiafa is with Instituto Argentino de Radioastronom\'{i}a (IAR) - CCT La Plata, CONICET CC5 (1894) V. Elisa, ARGENTINA and Facultad de Ingenier\'{i}a - UBA, Buenos Aires, ARGENTINA.\protect\\  
E-mail: ccaiafa@fi.uba.ar
\IEEEcompsocthanksitem Andrzej~Cichocki is with Lab. for Advanced Brain Signal Processing (LABSP), Brain Science Institute, RIKEN, 2-1 Hirosawa, Wako, 351-0198, JAPAN and Systems Research Institute, PAS, Warsaw, POLAND.

This paper has supplementary downloadable material available at http://ieeexplore.ieee.org., provided by the author. The material includes the Matlab codes needed to reproduce the simulation results. Contact ccaiafa@fi.uba.ar for further questions about this work.}
}

%
%

\markboth{Submitted to IEEE Transactions on Signal Processing}%
{Caiafa \MakeLowercase{\textit{et al.}}: Stable, Robust  and Super Fast Reconstruction of Tensors Using Multi-Way Projections}
%



\IEEEtitleabstractindextext{%
\begin{abstract}
In the framework of multidimensional Compressed Sensing (CS), we introduce an analytical reconstruction formula that allows one to recover an $N$th-order data tensor $\tX\in{\R^{I_1\times I_2\times \cdots \times I_N}}$ from a reduced set of multi-way compressive measurements by exploiting its low multilinear-rank structure. Moreover, we show that, an interesting property of multi-way measurements allows us to build the reconstruction based on compressive linear measurements taken only in two selected modes, independently of the tensor order $N$. In addition, it is proved that, in the matrix case and in a particular case with $3$rd-order tensors where the same 2D sensor operator is applied to all mode-3 slices, the proposed reconstruction $\tX_\tau$ is stable in the sense that the approximation error is comparable to the one provided by the best low-multilinear-rank approximation, where $\tau$ is a threshold parameter that controls the approximation error. Through the analysis of the upper bound of the approximation error we show that, in the 2D case, an optimal value for the threshold parameter $\tau=\tau_0 > 0$ exists, which is confirmed by our simulation results. On the other hand, our experiments on 3D datasets show that very good reconstructions are obtained using $\tau=0$, which means that this parameter does not need to be tuned. Our extensive simulation results demonstrate the stability and robustness of the method when it is applied to real-world 2D and 3D signals. A comparison with state-of-the-arts sparsity based CS methods specialized for multidimensional signals is also included. A very attractive characteristic of the proposed method is that it provides a direct computation, i.e. it is non-iterative in contrast to all existing sparsity based CS algorithms, thus providing super fast computations, even for large datasets. 
\end{abstract}

\begin{IEEEkeywords}
Compressed Sensing (CS), Kronecker-CS, Low-rank tensor approximation, Multi-way analysis, Tucker model.
\end{IEEEkeywords}}

\maketitle

\IEEEdisplaynontitleabstractindextext

%
\IEEEpeerreviewmaketitle

\section{Introduction}
\label{sec:intro}
\IEEEPARstart{D}{uring} the last years there has been an increased interest in \textit{Compressed Sensing} (CS), whose aim is the reconstruction of signals based on a set of measurements that is much smaller than the original signal size. Thus, instead to acquire a potentially large signal and compress it, CS suggests that signals can be compressively sampled thus reducing the amount of measurements. More specifically, in standard CS  \cite{CandesTao2006,Donoho:2006p424}, a signal $\x \in{\R^{n}}$, which is assumed unavailable, is reconstructed from a reduced set of $m$ linear projections ($m\ll n$) $\w = \Ph \x \in{\R^m}$, where the sensing matrix $\Ph\in{\R^{m\times n}}$ is typically random or composed by few selected rows of the Fourier transform matrix \cite{Candes:2006p425}. In order to make the CS problem solvable, it is necessary to exploit {\it a priori} information about the signal of interest $\x$ by imposing some constraints. For example, it is widely assumed that the signal $\x$ is compressible by decomposing it in a known Wavelet basis (\textit{dictionary}). In other words, it is assumed that every signal admits a \textit{sparse} representation on a given dictionary, i.e. combining only few elements, called \textit{atoms}. Under the scope of these \textit{sparse models}, many efficient CS algorithms were developed in order to reconstruct signals from compressive measurements which involve iterative refinements of the solution by means of \textit{Greedy algorithms} or by minimizing the $\ell_1$-norm of the solution (see \cite{Marvasti:2012gs} for an up to date summary of algorithms). These algorithms have found many applications in diverse fields such as in medical imaging, surveillance, machine learning, etc \cite{Eldar:2012wf}.

\subsection{Multidimensional CS}
Most of the development of CS was focused on problems involving 1D signal or 2D image data encoded in vectors. However, many important applications involve higher dimensional signals or \textit{tensors}. Some data sources are readily generated as tensors such as hyperspectral images,  videos, 3D light field displays \cite{Tensor-dispalys1}, Magnetic Resonance Imaging (MRI) \cite{Yu:2014jw}, etc., in other cases, tensors can be synthetically created by a rearrangement of lower dimensional data structures or by mathematical construction \cite{Cichocki:2013vf}. In some applications, like in materials science \cite{Nico:2013ux} or in scientific computation \cite{OseledetsTT09}, the exponential increase in memory and time requirements when the number of dimensions increases, makes impossible to work with full datasets and models with few parameters must be used. Such models are referred as tensor decompositions, which can be obtained by making few inspections of the full-datasets or by taking compressive measurements.

Recently, the Kronecker-CS model \cite{Duarte:cv} has been proposed in order to provide a practical implementation of CS for higher order tensors by exploiting their multidimensional structure. This model explicitly assumes that multidimensional signals have sparse representations using separable dictionaries, usually known as \textit{Kronecker dictionaries}. Kronecker bases are well known and widely used in image processing, for example, given a 3D $(I_1\times I_2\times I_3)$ image, its associated dictionary $\D\in{\R^{I_1I_2I_3\times I_1I_2I_3}}$ is $\D=\D_3\otimes \D_2\otimes \D_1$ where $\D_n\in{\R^{I_n\times I_n}}$ ($n=1,2,3$) are small dictionaries associated to columns (mode-1), rows (mode-2) and tubes (mode-3), respectively. Moreover, when working with multidimensional signals, the Kronecker structure also arises naturally in the physical implementation of the sensing devices since they can operate on different dimensions or modes of the signal, independently, through  separated sensing matrices \cite{Duarte:cv,August:2013ke}. This Kronecker structure of the sensing operator/dictionary is equivalent to apply the constrained Tucker model \cite{Caiafa:2012iv} and made possible to implement relatively fast and practical algorithms based on sparsity structures, e.g., on hyperspectral 3D images and video data through a vector $\ell_1$-norm minimization algorithm \cite{Duarte:cv}. 
More recently, greedy algorithms, especially designed to take advantage of the Kronecker structure and block sparsity of the representations, were proposed in \cite{Caiafa:2012iv} and applied to a variety of multidimensional signal processing problems such as in MRI, hyperspectral imaging and multidimensional inpainting \cite{Caiafa:2013jr}. Also, in \cite{Li:2013tr}, the authors developed generalized tensor compressed sensing algorithms by exploiting the Kronecker structure in a similar way as done in \cite{Caiafa:2012iv,Caiafa:2012vw} but using an $\ell_1$-minimization approach.

\subsection{Exploiting low-rank approximations instead of sparsity}
While sparsity is the ``working horse'' of standard CS, recently a new line of research has been proposed suggesting that, instead of using sparse models as {\it a priori} information about multidimensional signals, the low-rank approximation property could be exploited,  i.e. without any a priori knowledge about the possible bases or factor matrices for each mode (dictionaries). This idea was first explored in \cite{Candes:2010p386}, where matrices were reconstructed from its under-sampled measurements by solely assuming the existence of a low-rank approximation and by solving a convex optimization problem involving the matrix nuclear norm. These ideas have been extended to tensors, by considering different models for the measurements and using tensor low-rank approximations (based on the CANDECOMP/PARAFAC (CP) model) or low-multilinear-rank approximations (based on the Tucker model). For example, in \cite{JiLiu:bh}, tensor completion of visual data was analyzed by generalizing the minimization of matrix nuclear norm to the tensor case. In \cite{Acar:2011dja}, also the problem of estimating missing entries in tensors was considered by assuming that a low-rank CP model is fitted through a weighted least squares problem formulation.  In \cite{Golbabaee:2012uv}, hyperspectral images (3D tensors) are recovered from random linear projections of all channels by using a reconstruction algorithm that combines low-rank and join-sparse matrix recovery. In \cite{Rauhut:ta} the problem of reconstructing tensors having a low multilinear-rank Tucker model, based on a set of  linear measurements, was investigated and used for tensor denoising via an iterative multilinear-rank minimization method. In the context of optical-interferometric imaging, in \cite{2014MNRAS.437.2083A}, a method for recovering a supersymmetric rank-1 3D tensor from a set of multilinear measurements was proposed by using a convex linear program. In \cite{Sidiropoulos:kf}, the Kronecker sensing structure was used for tensor compression and a method involving a low-rank model fitting, followed by a per mode $\ell_0/\ell_1$ decompression, was proposed in order to recover a low-rank CP tensor with sparse factors. We refer to Table \ref{Table1} for a brief summary of recent approaches to CS involving tensor datasets, where the differences and similarities among the methods are highlighted.

\minrowclearance 1ex
\begin{table*}
\caption{Summary and comparison of available approaches to CS for tensor datasets } 
\scalebox{1.0}{
\centering
 {
    \begin{tabular*}{\linewidth}[t]{@{\extracolsep{\fill}}|p{2.3cm}|  p{1cm} |  p{4.2cm}|p{3.1cm}| p{3.6cm}| } \hline
{\bf Articles} & {\bf Tensor order} & {\bf Data Model} & {\bf Measurements Model} & {\bf Algorithm type}\\
\hline
{Duarte et al \cite{Duarte:cv}} & {$N \ge 2$} & {Sparsity, Kronecker dictionaries} & {Multilinear} & {$\ell_1$-norm minimization}\\
\hline
{Caiafa et al \cite{Caiafa:2012iv,Caiafa:2013jr}} & {$N \ge 2$} & {Sparsity, Block-sparsity, Kronecker dictionaries} & {Multilinear} & {Greedy}\\
\hline
{Li et al \cite{Li:2013tr}} & {$N \ge 2$} & {Sparsity, Kronecker dictionaries} & {Multilinear} & {$\ell_1$-norm minimization}\\
\hline
{Candes et al \cite{Candes:2010p386}} & {$N = 2$} & {Low-rank matrix} & {Linear (missing entries)} & {Nuclear norm minimization}\\
\hline
{Liu et al  \cite{JiLiu:bh}} & {$N = 3$} & {Low-rank (CP model)} & {Linear (missing entries)} & {Tensor nuclear norm minimization}\\
\hline
{Acar et al \cite{Acar:2011dja}} & {$N = 3$} & {Low-rank (CP model)} & {Linear (missing entries)} & {Weighted Least Squares}\\
\hline
{Golbabaee et al \cite{Golbabaee:2012uv}} & {$N = 2$} & {Joint-sparsity, Low-rank matrices} & {Linear} & {Nuclear norm, $\ell_{2,1}$ mixed norm minimization.}\\
\hline
{Rauhut et al \cite{Rauhut:ta} } & {$N \ge 3$} & {Low multilinear-rank (Tucker model)} & {Linear} & {Iterative Hard Thresholding}\\
\hline
{Aur\'{i}a et al \cite{2014MNRAS.437.2083A} } & {$N = 3$} & {Rank-1 Tensor} & {Multilinear} & {Convex programming}\\
\hline
{Sidiropoulos et al \cite{Sidiropoulos:kf} } & {$N = 3$} & {Low-rank Sparse CP model} & {Multilinear} & {$\ell_0/\ell_1$-norm  decompression}\\
\hline
{Current article } & {$N \ge 2$} & {Low multilinear-rank (Tucker model)} & {Multilinear} & {Non-iterative (direct) reconstruction}\\
\hline

    \end{tabular*}
   }}
\label{Table1}
\end{table*}

In this work, we extend the ideas and results of our recent conference paper \cite{Caiafa:2013uu}, providing a direct (i.e., analytical) reconstruction formula that allows us to recover a tensor from a set of multilinear projections that are obtained by multiplying the data tensor by a different sensing matrix in each mode. This model comes into scene naturally in many potential applications, for example, in the case of sensing 2D or 3D images by means of a separable operator as developed in \cite{Robucci:2010cw,Rivenson:2009p285,Duarte:cv,August:2013ke}, i.e., by taking compressive measurements of columns, rows, etc. separately, imposing a Kronecker structure on the sensing operator. The key assumption is that our multidimensional signal is well approximated by a low multilinear-rank Tucker model which is realistic for many structured datasets, specially in the case of multidimensional images. We formulate our multidimensional CS model in a general setting for $N$-th order tensors and provide theoretical stability analysis and robustness evidence for $N=2$ and a very important particular case with $N=3$. 

A particularly distinctive and attracting feature of the proposed reconstruction method is that, unlike all other methods listed in Table \ref{Table1}, our method is non-iterative. We believe that the present mathematical model could be fully exploited by the next generation of multidimensional compressive sensors for very large datasets. Through extensive simulations on real-world datasets, we also illustrate the relevance of our results in hyperspectral compressive imaging for which the technology is already available \cite{Duarte:cv,Robucci:2010cw,August:2013ke}.

\subsection{Paper organization}
This paper is organized as follows: in Section \ref{sec:Intro}, tensor notation, definitions and basic results used throughout the paper, are introduced; in Section \ref{sec:exact}, the reconstruction formula is introduced for the ideal case when the tensor of interest admits an exact low multilinear-rank representation; in Section \ref{sec:Lowrank}, the effect of a more realistic model for signals is analyzed and a modified reconstruction formula is proposed in order to guarantee stable reconstructions; in Section \ref{sec:Simulations}, several numerical results based on 2D and 3D real-world signals are provided, validating our theoretical results and evaluating the stability and robustness of our proposed reconstruction scheme. The performance is evaluated in terms of computational time, quality of reconstructions and variance of the results over Monte Carlo simulations (robustness). Finally, in section \ref{sec:Conclusions}, the main conclusions of the present work are outlined.

\section{Notation, definitions and preliminary results} \label{sec:Intro}

\subsection{Tensor notation and operations}
Tensors (multi-way arrays) are denoted by underlined boldface capital letters, e.g. $\tX\in{\R^{I_1\times I_2\times \cdots \times I_N}}$ is an $N$-th order tensor of real numbers. Matrices (2D arrays) are denoted by bold uppercase letters and vectors by boldface lower-case letters, e.g.  $\X\in{\R^{I_1\times I_2}}$ and $\x \in{\R^I}$ are a matrix and a vector, respectively. The element $(i_1,i_2,\dots,i_N)$ of a tensor is referred as $x_{i_1i_2\dots i_N}$. The Frobenius norm is defined by $\|\tX\|_F=\sqrt{\sum_{i_1}\cdots \sum_{i_N}x_{i_1i_2\dots i_N}^2}$. The spectral norm of a matrix $\A$ is denoted by $\|\A\|$ corresponding to its largest singular value.

Given a tensor $\tX\in{\R^{I_1\times I_2\times \cdots \times I_N}}$, its mode-$n$ fibers are the vectors obtained by fixing all indices except $i_n$, which correspond to columns ($n=1$), rows ($n=2$), and so on. Mode-$n$ unfolding of a tensor $\tX\in{\R^{I_1\times I_2\times \cdots \times I_N}}$ yields a matrix $\X_{(n)}\in{\R^{I_n\times \bar{I}_n}}$ ($\bar{I}_n=\prod_{m\ne n} I_m$) whose columns are the corresponding mode-$n$ fibers arranged in a particular order, to be more precise, tensor element $(i_1,i_2,\dots,i_N)$ maps to matrix element $(i_n,j)$, where $j=1+\sum_{k\neq n}(i_k-1)J_k$ with $J_k=\prod_{m\neq n}^{k-1}I_m$ \cite{Kolda:2009vq}. 

Given a multidimensional signal (tensor) $\tX\in{\R^{I_1\times I_2\times \cdots \times I_N}}$ and a matrix $\Ph \in{\R^{J\times I_n}}$ the mode-$n$ tensor by matrix product $\tY=\tX \times_n \Ph \in{\R^{I_1\times \cdots \times I_{n-1}\times J \times I_{n+1}\times \cdots \times I_N}}$ is defined by: 
\begin{equation}\label{tenbymat_prod}
y_{i_1 \cdots i_{n-1} j i_{n+1} \cdots i_N} = \sum_{i_n=1}^{I_n} x_{i_1\cdots i_n \cdots i_N}\phi_{ji_n},
\end{equation}
with $i_k=1,2,...,I_k$ ($k\neq n$) and $j=1,2,...,J$. It should be noted that this corresponds to the product of matrix $\Ph$ by each one of the mode-$n$ fibers of $\tX$ since $\Y_{(n)} = \Ph \X_{(n)}$.

\subsection{Tucker model and multilinear-rank}
The {\it Tucker decomposition model} \cite{DeLathauwer2000} provides a generalization of the low-rank approximation of matrices to the case of tensors, i.e. for a given tensor $\tX\in{\R^{I_1\times I_2\times \cdots \times I_N}}$, we have $\tX = \tX_0 + \tE$, where $\tE$ is an error tensor and the multilinear-rank-($R_1,R_2,\dots,R_N$) tensor approximation $\tX_0$ is defined as follows (Tucker model):
\begin{equation}\label{Tucker}
\tX_0=\tG \times_1 \A_1 \times_2 \cdots \times_N \A_N,
\end{equation}
with a \emph{core tensor} $\tG\in{\R^{R_1\times  R_2\times \cdots \times R_N}}$ and \emph{factor matrices} $\A_n \in{\R^{I_n\times R_n}}$ (typically $R_n\ll I_n$). A data tensor $\tX\in{\R^{I_1\times I_2\times \cdots\times I_N}}$ is said to have \textit{multilinear-rank}-$(R_1,R_2,\dots,R_N)$ if such a decomposition is exact for a set of minimal values $(R_1,R_2,\dots,R_N)$, i.e. $\tX = \tX_0$. We say that a tensor  $\tG\in{\R^{R_1\times  R_2\times \cdots \times R_N}}$ is \textit{full-rank} if all its unfolded matrices are full-rank, i.e., $\rank{(\G_{(n)})}= R_n$, $\forall n$.
A particularly interesting case of the Tucker model is when factor matrices $\A_n =\U_n\in{\R^{I_n\times R_n}}$ are orthogonal and chosen as the truncated matrices of left singular vectors associated with the unfolding matrices $\X_{(n)}=\U_n\mathbf{\Sigma}_n\V_n^T$. In this case, we obtain the so called truncated \textit{Higher Order Singular Value Decomposition} (HOSVD) \cite{DeLathauwer2000}. 
It is noted that, in the matrix case, the truncated SVD provides the best low rank approximation having orthogonal factors and a diagonal core matrix.

\subsection{Multi-way Projections}
While in classical 1D CS, the set of compressive measurements is obtained by a linear projection, i.e. by multiplying the vector signal by a sensing matrix, in the tensor case, we can exploit its multi-way structure and use devices that provide compressive measurements by multiplying an $N$th-order tensor by sensing matrices in several modes, similarly or identically to the Kronecker-CS setting \cite{Duarte:cv}. According to the definition of the mode-$n$ product in eqn. (\ref{tenbymat_prod}), multiplying a data tensor by a sensing matrix in the mode-$n$ corresponds to apply a projection to every mode-$n$ fiber. For example, a 2D signal $\X \in{\R^{I_1\times I_2}}$ can be compressively sensed by using two sensing matrices, $\Ph_1\in{\R^{R_1\times I_1}}$ and $\Ph_2\in{\R^{R_2\times I_2}}$ for mode-1 and mode-2, respectively, i.e. $\W= \X \times_1 \Ph_1 \times_2 \Ph_2 \in{\R^{R_1\times R_2}}$ or, equivalently, $\W= \Ph_1 \X  \Ph_2^T$, or  $\w=  (\Ph_2 \otimes \Ph_1)^T\x$, where $\w\in{\R^{R_1R_2}}$ and $\x\in{\R^{I_1I_2}}$ are the vectorized versions of matrices $\W$ and $\X$, respectively. Thus, the objective of Kronecker-CS is to recover the signal $\X$ from the measured data matrix $\W$. 

\begin{figure}[!t]
 \centering
 \centerline{\includegraphics[width=12cm]{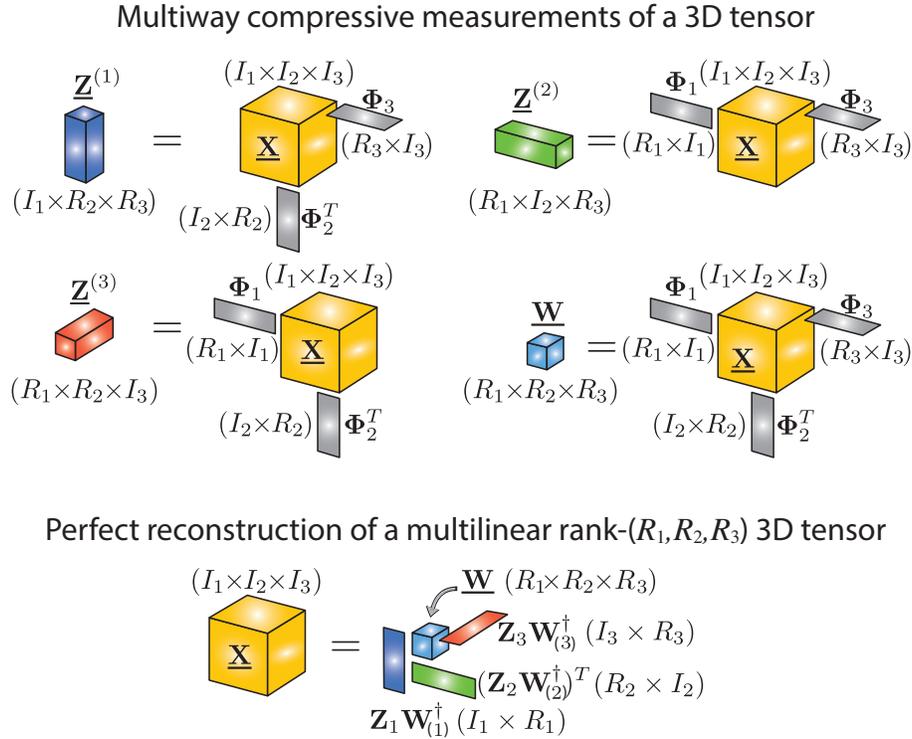}}
 \caption{\footnotesize{Multi-way measurements and the reconstruction model for a low multilinear-rank 3D tensor.}}
   \label{FigModel}
\end{figure}

In this paper, we assume that the following set of compressive multi-way measurements $\underline{\Z}^{(n)}\in{\R^{R_1\times \cdots \times R_{n-1} \times I_n \times R_{n+1} \times \cdots \times R_N}}$ ($n=1,2,\dots,N$), are available:
\begin{equation} \label{multimeas}
\underline{\Z}^{(n)} = \tX \times_1 \Ph_1 \times_2 \cdots\times_{n-1}\Ph_{n-1}\times_{n+1}\Ph_{n+1}\times_{n+2} \cdots \times_{N}\Ph_{N},
\end{equation}
where $\Ph_n\in{\R^{R_n \times I_n}}$ ($R_n\ll I_n$) are the corresponding mode-$n$ sensing matrices. Note that eqn. (\ref{multimeas}) indicates that the original tensor is multiplied by the set of sensing matrices in all modes except in mode-$n$ (see Fig. \ref{FigModel} (top)). We also assume available the following core tensor $\underline{\W}\in{\R^{R_1\times R_2\times \cdots \times R_N}}$:
\begin{equation} \label{core}
\underline{\W} = \tX \times_1 \Ph_1 \times_2 \cdots \times_{N}\Ph_{N},
\end{equation}
which, in fact, is redundant since it can be computed from $\tZ^{(n)}$ and $\Ph_{n}$ taking into account that $\tW = \tZ^{(n)} \times_n \Ph_{n}$ for any $n$.

Most of state-of-the-art tensor reconstruction algorithms based on Kronecker-CS \cite{Duarte:cv,Caiafa:2012iv,Caiafa:2013jr,2014MNRAS.437.2083A,Sidiropoulos:kf} assume that the only available measurement tensor is $\underline{\mathbf{W}}$, i.e. the product of the original dataset by the sensing matrices $\mathbf{\Phi}_n$ (n=1,2,3) in all modes simultaneously. On the other hand, our present method requires to have the set of tensor measurements obtained by multiplying the datasets by all the sensing matrices except one, i.e. tensors $\underline{\mathbf{Z}}^{(n)}$ (n=1,2,3),  not necessarily involving a larger amount of measurements (see experimental comparison in Section \ref{sec:exp_comparison}). However, it is not difficult to see that already existing hardware implementations of CS imaging systems can be easily adapted in order to provide the kind of measurements required by our method. For example, in the 2D case, our method proposes to collect measurements on columns using a common sensing matrix $\mathbf{\Phi}_1$, and rows using $\mathbf{\Phi}_2$. By using the same ideas of the single-pixel camera developed in \cite{Duarte:er} and used in \cite{Duarte:cv}, our sensing operator can be obtained by using, for example, a linear array of DMDs (Digital Microarray Devices) with random orientations to sense columns and another linear array of DMDs to sense rows. On the other hand, it is also interesting to note that in \cite{Robucci:2010cw}, a different hardware implementation was used to provide real Kronecker-CS, i.e., by applying different sensing matrices to columns and rows which can be used to provide the measurements required by our method. It is also interesting to note that, recently, in \cite{August:2013ke}, a new hardware implementation has been proposed that allows to employ Kronecker-CS with separable sensing operators in space and in spectral domains which could be also used to provide the measurements required by our method.

In some 3D applications, the mode-$3$ sensing matrix is the identity matrix, i.e. $\Ph_{3}=\I \in{\R^{I_3 \times I_3}}$, because the same sensing operator $(\Ph_2\otimes \Ph_1)$ is applied to each frontal slice of the tensor. This is the case, for instance, in hyperspectral compressive imaging, where each frequency band (a frontal slice) is sensed by applying a different selective filter \cite{Duarte:cv,2014ISPM...31..116W}. This mathematical model is also valid for video sequences, where each frontal slice of the tensor corresponds to a snapshot taken at a given time.

\subsection{The Truncated Moore-Penrose Pseudo-inverse} \label{sec:MoorePenrose}
It is well known that, by using the Moore-Penrose (MP) pseudo-inverse of an ill-conditioned matrix $\A\in{\R^{I_1\times I_2}}$, an unstable behavior is produced since the norm $\|\A^\dagger\|$ could be extremely large when the smallest singular value $\sigma_{\min{(I_1,I_2)}}$ is close to zero. In order to avoid this problem, the \textit{truncated MP pseudo-inverse} $\W^{\ast_\tau}$ can be used as a regularization technique, which is defined as follows \cite{Hansen:1998vo}:
\begin{equation}
\W^{\ast_\tau} = \V \S^{\ast_\tau} \U^T,
\end{equation}
with entries of the diagonal matrix $\S^{\ast_\tau} \in{\R^{I_2 \times I_1}}$ defined as follows:
\begin{equation}
\sigma^{\ast}_i  =  \left\{
  \begin{array}{l l}
    1/\sigma_i, & \quad \text{if $\sigma_i >\tau$}\\
   0, & \quad \text{if $\sigma_i \le \tau$},
  \end{array} \right.   
\end{equation}
where $\tau$ is a free threshold parameter (see next sections). It is noted that $\W^{\ast_\tau} \rightarrow \W^\dagger$ as $\tau \rightarrow 0$ and $\|\W^{\ast_\tau}\| \le 1/\max{(\tau,\sigma_R)}$. Also, the following properties are easily verified: 
\begin{equation}\label{PropI}
\W \W^{\ast_\tau} \W = \W + \H,
\end{equation}
with $\|\H\| \le \tau$ if $\tau > \sigma_R$, and $\H=\0$ if $\tau \le \sigma_R$;
\begin{eqnarray}
\W^{\ast_\tau} \W \W^{\ast_\tau} = \W^{\ast_\tau}, \label{PropII} \mbox{ and} \\
\|\W \W^{\ast_\tau}\| = \| \W^{\ast_\tau} \W \|=1. \label{PropIII}
\end{eqnarray}

The following lemma, provides a generalization of the property (\ref{PropI}) to the $3$rd order tensor case.
\begin{lemma} \label{lemmaTensor}
For a given tensor $\tW \in{\R^{R_1\times R_2 \times R_3}}$ with smallest singular values in each mode $\sigma_{R_1}$, $\sigma_{R_2}$ and $\sigma_{R_3}$, respectively, the following property holds:
\begin{equation}\label{lefthandLemma}
\tW \times_1 \W_{(1)}\W_{(1)}^{\ast_\tau} \times_2 \W_{(2)}\W_{(2)}^{\ast_\tau} \times_3 \W_{(3)}\W_{(3)}^{\ast_\tau} = \tW + \tH,
\end{equation}
with 
\begin{eqnarray}\nonumber
\tH = \t0,& \quad &\text{if $\tau \le \underline{\sigma}$}, \\
\|\tH\|_F \le &(\sqrt{R_1}+\sqrt{R_2}+\sqrt{R_3})\tau, \quad &\text{if $\tau > \overline{\sigma}$} \nonumber
\end{eqnarray}
where $\overline{\sigma} =\max{(\sigma_{R_n})}$ and $\underline{\sigma} =\min{(\sigma_{R_n})}$ ($n=1,2,3$).
\begin{proof}
See proof in the Appendix A.
\end{proof}
\label{propTensor}
\end{lemma} 

\section{Exact low multilinear-rank tensor recovery}
\label{sec:exact}
The following theorem provides an explicit reconstruction formula, as illustrated in Fig. \ref{FigModel} (bottom), and states the conditions under which the original tensor can be exactly recovered from the set of multi-way measurements $\tZ^{(n)}$ ($n=1,2,\dots,N$) and $\tW$, defined in equations (\ref{multimeas}) and (\ref{core}), respectively. 
\begin{theorem}[\bf Low multilinear-rank case] If tensor $\tX\in{\R^{I_1\times I_2\times \cdots \times I_N}}$ has multilinear-rank-$(R_1,R_2\dots,R_N)$ and sensing matrices $\Ph_n\in{\R^{R_n \times I_n}}$ are such that the tensor $\tW = \tX \times_1 \Ph_1 \cdots \times_N \Ph_N \in{\R^{R_1\times R_2\times \cdots \times R_N}}$ is full-rank, then the following reconstruction formula is exact, i.e. $\hat{\tX} = \tX$:
\begin{equation}\label{reconstexact}
\hat{\tX} = \tW \times_1 \Z_1\W_{(1)}^\dagger \times_2 \cdots \times_N \Z_N\W_{(N)}^\dagger,
\end{equation}
where ``$^\dagger$'' stands for the MP pseudo-inverse of a matrix and $\Z_n \equiv (\underline{\Z}^{(n)})_{(n)}\in{\R^{I_n\times \bar{R}_n}}$, with $\bar{R}_n=\prod_{m\ne n} R_m$.
\label{theoremExact}

\begin{proof}
Let us consider the exact HOSVD decomposition $\tX = \tGamma \times_1\U_1 \times_2 \cdots \times_N\U_N$, with core tensor $\tGamma \in{\R^{R_1\times R_2\times \cdots \times R_N}}$ and orthogonal factors $\U_n \in{\R^{I_n\times R_n}}$, which exists because it is assumed that tensor $\tX$ has multilinear-rank-($R_1,R_2,\dots,R_N$). We consider, for convenience, a change of bases such that the new factors $\A_n$ satisfy $\Ph_n \A_n=\I \in{\R^{R_n\times R_n}}$. We can do this by defining the new set of factors as follows: 
\begin{equation}\label{basischange}
\A_n = \U_n (\Ph_n \U_n)^{-1},
\end{equation} 
thus, we have $\tX = \tG \times_1\A_1 \times_2 \cdots \times_N\A_N$ where $\tG = \tX \times_1 \A_1^{\dagger} \times_2 \cdots \times_N \A_N^{\dagger} = \tW$. Note also that, with these new bases, the multi-way measurements are now simplified to $\underline{\Z}^{(n)} = \tW \times_n \A_n$ or, equivalently, $\Z_n = \A_n\W_{(n)}$. Taking into account that $\W_{(1)}\W_{(1)}^\dagger \W_{(1)} = \W_{(1)}$, the mode-$1$ unfolded version of eqn. (\ref{reconstexact}) is:
\begin{equation}
\hat{\X}_{(1)} = \A_1 \W_{(1)} \left( \Z_N\W_{(N)}^\dagger \otimes \cdots \otimes \Z_2\W_{(2)}^\dagger \right)^T,
\end{equation}
which, can be written in terms of its associated mode-$2$ unfolding matrix as:
\begin{equation}\nonumber
\hat{\X}_{(2)} = \Z_2 \W_{(2)}^\dagger \W_{(2)}\left( \Z_N\W_{(N)}^\dagger \otimes \cdots \otimes \Z_3\W_{(3)}^\dagger \otimes \A_1\right)^T. 
\end{equation}
Now, by substituting $\Z_2 = \A_2\W_{(2)}$ in the previous equation and, by repeating this process for all modes $n=3,4,\dots,N$, we finally arrive at:
\begin{equation}
\hat{\X}_{(N)} = \A_N \W_{(N)}\left( \A_{N-1} \otimes \cdots \otimes \A_1\right)^T,
\end{equation}
which proves that $\hat{\tX} =  \tX$, since $\tW=\tG$.
\end{proof}
\end{theorem}

It is noted that, using a simplified notation, the Tucker model of a tensor with size $I\times I \times I$ having multilinear rank $(R,R,R)$, requires $R^3 + 3IR$ parameters and the suggested reconstruction needs no more than $3IR^2$ measurements corresponding to tensors $\tZ^{(n)}$, $n=1,2,3$. This is larger than the number of parameters of the associated Tucker model, however, if $R \ll I$ we see that the number of measurements is approximately $R$ times the number of parameters, which is in general much smaller that the total number of entries $I^3$.


\subsection{Multi-way Measurements Via Linear Projections Applied to Only Two Selected Modes}

It is interesting to note that all multi-way measurements defined in eqn. (\ref{multimeas}) ($n=1,2,\dots,N$) and eqn. (\ref{core}), can be computed from linear measurements taken only in two selected modes out of $N>$. To show this, suppose that we have at our disposal the linear measurements in modes $m_1$ and $m_2$ given by: $\Y_m = \Ph_m \X_{(m)}$, with $m=m_1,m_2$; then, it is easy to see that the mode-$m$ unfolding matrix of each multi-way measurement $\underline{\Z}^{(n)}$, $(n\neq m)$, can be written as follows:
\begin{eqnarray} \nonumber
(\underline{\Z}^{(n)})_{(m)} &=& \Y_{m} (\Ph_{N}^T \otimes  \cdots \otimes \Ph_{m+1}^T \otimes \Ph_{m-1}^T \otimes \cdots \\ 
& & \cdots \otimes \Ph_{n+1}^T \otimes \I \otimes \Ph_{n-1}^T  \otimes \cdots \otimes \Ph_{1}^T ). \nonumber   
\end{eqnarray}
For example, in the important particular case of hyperspectral images, sometimes it is easier to take compressive measurements of columns (mode-1) and rows (mode-2) of a 3rd order tensor $\tX\in{\R^{I_1\times I_2\times I_3}}$by using the corresponding matrices $\Ph_1\in{\R^{R_1\times I_1}}$ and $\Ph_1\in{\R^{R_2\times I_2}}$ as follows:
\begin{equation}\label{linproj}
\Y_1 = \Ph_1 \X_{(1)} \mbox{  and  } \Y_2 = \Ph_2 \X_{(2)}.
\end{equation}
It is easy to see that the multi-way measurements required by our method can be obtained from these two set of linear projections, for example, noting that: $(\underline{\Z}^{(1)})_{(2)} =  \Y_2 ( \Ph_3^T \otimes \I)$, $(\underline{\Z}^{(2)})_{(1)} =  \Y_1 (\Ph_3^T \otimes \I)$, $(\underline{\Z}^{(3)})_{(1)} =  \Y_1 (\I \otimes \Ph_2^T)$,
and that tensor $\tW$ can be obtained using that $\tW=\tZ^{(n)}\times_n \Ph_n$ (for any $n$). In this case, matrix $\Ph_3\in{\R^{R_3\times I_3}}$ is not a real sensing matrix and its rank $R_3$ must be chosen in order to capture the numerical rank of the dataset in its 3rd mode, i.e. $\X_{(3)}$. In Algorithm \ref{alg3D}, the steps to reconstruct a tensor $\tX \in \R^{I_1\times I_2\times I_3}$ from the linear projections defined in equation (\ref{linproj}) are summarized.

\begin{algorithm}
{\footnotesize
\caption{: Reconstruction of tensor $\tX$ from linear projections taken in mode-1 and mode-2 (as defined in eqn. (\ref{linproj}))}
\label{alg3D}
\begin{algorithmic}[1]
\REQUIRE Linear projections $\Y_1\in{\R^{R_1\times I_2I_3}}$, $\Y_2\in{\R^{R_2\times I_1I_3}}$, sensing matrices $\Ph_1,\Ph_2,\Ph_3\in{R^{R_n\times I_n}}$  ($R_n\ll I_n$)
\ENSURE Tensor reconstruction $\hat{\tX}\in{\R^{I_1\times I_2\times I_3}}$
\STATE $(\underline{\Z}^{(1)})_{(2)} =  \Y_2 ( \Ph_3^T \otimes \I)$; Mode-2 unfolding of $\underline{\Z}^{(1)}$ 
\STATE $\underline{\Z}^{(1)} = \mbox{tensorize}((\underline{\Z}^{(1)})_{(2)})$; Tensorization 
\STATE $(\underline{\Z}^{(2)})_{(1)} =  \Y_1 ( \Ph_3^T \otimes \I)$; Mode-1 unfolding of $\underline{\Z}^{(2)}$ 
\STATE $\underline{\Z}^{(2)} = \mbox{tensorize}((\underline{\Z}^{(2)})_{(1)})$; Tensorization 
\STATE $(\underline{\Z}^{(3)})_{(1)} =  \Y_1 (\I \otimes \Ph_2^T)$; Mode-1 unfolding\footnotemark of $\underline{\Z}^{(3)}$ 
\STATE $\underline{\Z}^{(3)} = \mbox{tensorize}((\underline{\Z}^{(3)})_{(1)})$; Tensorization 
\STATE $\Z_n = (\underline{\Z}^{(n)})_{(n)}$; Mode-$n$ unfolding ($n=1,2,3$) 
\STATE $\tW = \underline{\Z}^{(n)} \times_n \Ph_n$; for any $n=1,2,$ or $3$
\STATE Compute pseudo inverses of unfolding matrices $\W_{(n)}$ ($n=1,2,3$)
\STATE $\hat{\tX} = \tW \times_1 \Z_1\W_{(1)}^\dagger \times_2 \Z_2\W_{(2)}^\dagger \times_3 \Z_3\W_{(3)}^\dagger$,
\RETURN $\hat{\tX}$
\end{algorithmic}
}
\end{algorithm}
\footnotetext{It is noted that the mode-2 unfolding can be used instead as follows: $(\underline{\Z}^{(3)})_{(2)} =  \Y_2 (\I \otimes \Ph_1^T)\in{\R^{R_2\times I_3R_1}}$.}

\section{Stable Reconstructions of Approximately Low multilinear-Rank Tensors} \label{sec:Lowrank}
The reconstruction formula of eqn. (\ref{reconstexact}) is exact for the case where the tensor has multilinear-rank-($R_1,R_2,\dots,R_N$). However, in real world applications, signals usually have not low multilinear-rank, but they admit good low multilinear-rank approximations. Thus, a more realistic model for a signal should be $\tX = \tX_0 + \tE$ where $\tX_0$ has exact multilinear-rank-($R_1,R_2,\dots,R_N$) and $ \tE$ is a tensor error with sufficiently small norm, i.e. $\|\tE\|_F \le \epsilon$. 

In this case, one may ask about the optimal choice for sensing matrices $\Ph_n \in{\R^{R_n \times I_n}} $ in order to provide the best low multilinear-rank reconstruction. In the matrix case ($N=2$), it is not difficult to see that, if sensing matrices are constructed using the first $R$ singular vectors in each mode, then the obtained reconstruction is optimal (best low-rank approximation). To be more specific, given the SVD of the data matrix $\X = \U \mathbf{\Lambda} \V^T\in{\R^{I_1\times I_2}}$ with $\U = \bigl(
\begin{matrix} 
\U_1& \U_2
\end{matrix} 
\bigr)$, 
$\V =\bigl(
\begin{matrix} 
\V_1& \V_2
\end{matrix} 
\bigr)$ and 
$\mathbf{\Lambda} = \bigl(
\begin{smallmatrix} 
\mathbf{\Lambda}_1& \mathbf{0}\\ 
\mathbf{0} & \mathbf{\Lambda}_2
\end{smallmatrix} 
\bigr)$
where $\U_1\in{\R^{I_1\times R}}$ and $\V_1\in{\R^{I_2\times R}}$ are the first $R$ left and right singular values, respectively, and $\mathbf{\Lambda}_1$ is a diagonal matrix containing the first $R$ singular values in its main diagonal in decreasing order, if we define the sensing matrices as follows:  $\Ph_1 = \U_1^T \in{\R^{R\times I_1}}$, $\Ph_2 = \V_1^T\in{\R^{R\times I_2}}$, then the reconstruction is given by:
\begin{eqnarray}
\hat{\X} &=& \Z_1\W^\dagger \Z_2^T = (\X\Ph_2^T) (\Ph_1\X\Ph_2^T)^\dagger (\X^T\Ph_1^T) = 
(\X\V_2) (\U_1^T\X\V_2)^\dagger (\X^T\U_1)= \nonumber\\
&=& (\U_1\mathbf{\Lambda}_1)\mathbf{\Lambda}_1^\dagger (\mathbf{\Lambda}_1 \V_1^T) = \U_1\mathbf{\Lambda}_1\V_1^T = \X_0,
\end{eqnarray}
where $\X_0$ is, by definition the truncated SVD which is the best low rank approximation.

However, in practice we do not know the singular vectors because the original dataset is not available so we need to use sensing matrices that are independent from the dataset, for example, by generating them randomly which will give us sub-optimal reconstructions, i.e. $\|\tX - \hat{\tX}\|_F \ge \epsilon$. 

In particular, we say that a reconstruction method is stable if the obtained error $\|\tX - \hat{\tX}\|_F$ is comparable to the input error $\epsilon$, i.e. $\|\tX - \hat{\tX}\|_F \sim K\epsilon$ for some constant $K$. As we will show in this section, the formula given by eqn. (\ref{reconstexact}) may suffer from an unstable behavior, especially in the matrix case ($N=2$), i.e. generating large output errors even when the input error $\epsilon$ is small. We will show that we can solve this unstable behavior by using the truncated pseudo-inverse, as defined in Section \ref{sec:MoorePenrose}, instead of the exact MP pseudo-inverse. In other words, we define the {\it modified reconstruction} formula as follows:
\begin{equation}\label{modified_reconst}
\hat{\tX}_\tau = \tW \times_1 \Z_1\W_{(1)}^{\ast_\tau} \times_2 \cdots \times_N \Z_N\W_{(N)}^{\ast_\tau},
\end{equation}
where $\W_{(n)}^{\ast_\tau}$ is the truncated pseudo-inverse of matrix $\W_{(n)}$ (with the threshold parameter $\tau$). It is noted that, when $\tau \le \sigma_R(\W_{n})$ $\forall n$, eqn. (\ref{modified_reconst}) is equivalent to eqn. (\ref{reconstexact}). 

In the next sections we derive theoretical upper bounds for the reconstructions errors for $N=2$ (matrix case) and for a particular case of a 3D tensor ($N=3$), where the same 2D sensor is applied to every frontal slice. We also analyze under which conditions the parameter $\tau$ can be chosen in order to provide the minimum upper bound (optimal value $\tau_{opt}$).

\subsection{Error Upper Bound for the 2D case (N=2)}
The following theorem provides an upper bound for the reconstruction of eqn. (\ref{modified_reconst}) and shows that the error bound approaches to zero as $\epsilon \rightarrow 0$ if $\tau=0$ or $\tau \propto \epsilon$ (i.e., $\tau$ is proportional to the best approximation error $\epsilon$).
\begin{theorem} \label{theoremApprox2D}.
Let matrix $\X\in{\R^{I_1\times I_2}}$ be approximated by a rank-$R$ matrix $\X_0\in{\R^{I_1\times I_2}}$, i.e. $\X = \X_0 + \E$ where $\|\E\| \le \epsilon$ and, given sensing matrices $\Ph_1\in{\R^{R \times I_1}}$,$\Ph_2\in{\R^{R \times I_2}}$ such that $\W = \Ph_1 \X \Ph_2^T$ is full-rank, then the following error upper bound holds:
\begin{equation}\label{errorbound}
\|\X - \hat{\X}_\tau\|  \le  \left\{
  \begin{array}{l l}
    b \epsilon + c \frac{\epsilon^2}{\sigma_R}, & \quad \text{if $\tau \le \sigma_R$},\\
   a \tau + b \epsilon + c \frac{\epsilon^2}{\tau}, & \quad \text{if $\tau > \sigma_R$},
  \end{array} \right.   
\end{equation}
where $\sigma_R$ is the $R$-th singular value (smallest) of the matrix $\W$ and constants $a$, $b$ and $c$ are defined below.

\begin{proof}
Let $\X_0 = \U_1 \Gam \U_2^T$ be the truncated SVD of $\X_0$, and the factors $\A_1,\A_2 \in{\R^{I\times R}}$ are defined by eqn. (\ref{basischange}), then we obtain
\begin{equation}\label{result1}
\X_0 = \A_1 \G \A_2^T \mbox{   with  } \G = \W - \Ph_1\E\Ph_2^T.
\end{equation} 

The mode-$1$ measurement matrix is $\Z_1 = \X_{(1)}\Ph_2^T = \X_0\Ph_2^T + \E\Ph_2^T = \A_1\W_{(1)} + \F_1$, with $\F_1=(\I - \A_1\Ph_1)\E \Ph_2^T$ (where we assumed that $\Ph_n\A_n=\I$), and using a similar analysis for mode-$2$, we obtain that $\Z_2 =  \A_2\W_{(2)} + \F_2$, with $\F_2=(\I - \A_2\Ph_2)\E^T\Ph_1^T$.

Using the property (\ref{PropII}) of the truncated pseudo-inverse, the reconstructed matrix becomes:
\begin{equation}\label{reconst2D}
\hat{\X}_\tau =  \Z_1 \W^{\ast_\tau} \Z_2^T.
\end{equation}
Now, by inserting the expressions for $\Z_1$ and $\Z_2$ into eqn. (\ref{reconst2D}) we obtain:
\begin{equation}\label{Xdecomp}
\hat{\X}_\tau = \A_1\W\W^{\ast_\tau}\W \A_2^T + \A_1\W\W^{\ast_\tau}\F_2^T + \F_1\W^{\ast_\tau}\W \A_2^T + \F_1\W^{\ast_\tau} \F_2^T. 
\end{equation}

Assuming that $\W \W^{\ast_\tau} \W = \W + \H$ with $\|\H\| \le \tau$ if $\tau > \sigma_R$, and $\H=\0$, if $\tau \le \sigma_R$ (see Section \ref{sec:MoorePenrose}), and using $\X_0 = \X - \E$ and equation (\ref{result1}), we obtain:
\begin{equation}
\hat{\X}_\tau - \X = - \E +\A_1\Ph_1\E\Ph_2^T \A_2^T + \A_1\H\A_2^T + \A_1\W\W^{\ast_\tau}\F_2^T + \F_1\W^{\ast_\tau}\W \A_2^T + \F_1\W^{\ast_\tau}\F_2^T. 
\end{equation}
If we apply the spectral norm to the last matrix equation and by using that $\|\W^{\ast_\tau}\| = 1/\max{(\tau,\sigma_R)}$, we finally obtain
\begin{equation}
\|\X - \hat{\X}_\tau\|  \le  \left\{
  \begin{array}{l l}
    b \epsilon + c \frac{\epsilon^2}{\sigma_R}, & \quad \text{if $\tau \le \sigma_R$}\\
   a \tau + b \epsilon + c \frac{\epsilon^2}{\tau}, & \quad \text{if $\tau > \sigma_R$},
  \end{array} \right.   
\end{equation}
where constants are identified as follows:
\begin{align} 
a &=  \|\A_1\|\|\A_2\|, \\
b &= 1 + \|\A_1\Ph_1\| \|\A_2\Ph_2\| +  \|\A_1\|(1+\|\A_2\Ph_2 \|)\|\Ph_1\|  +  \|\A_2\|(1+\|\A_1\Ph_1 \|)\|\Ph_2\|, \\
c &= (1+\|\A_1\Ph_1\|)(1+\|\A_2\Ph_2\|)\|\Ph_1\| \|\Ph_2\|.
\end{align}


\end{proof}
\end{theorem}

\begin{figure}[!t]
 \centering
 \centerline{\includegraphics[width=12cm]{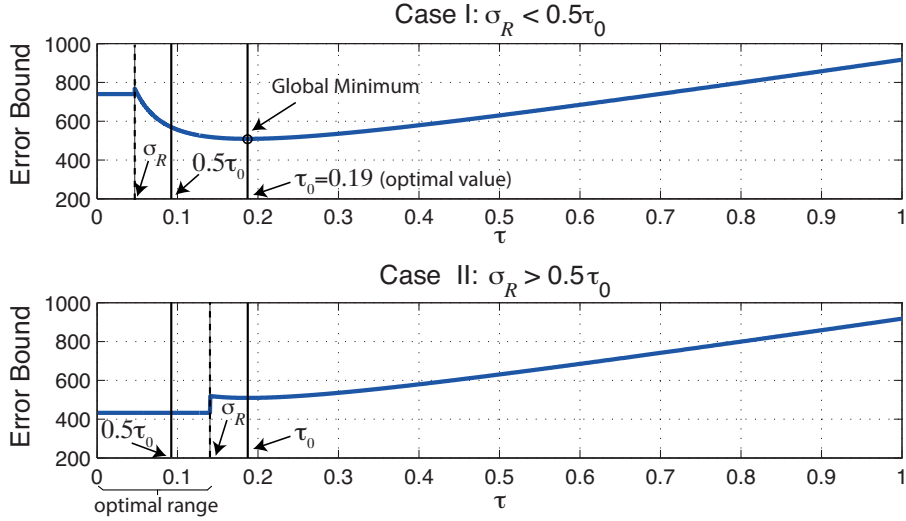}}
 \caption{\footnotesize{Matrix Case ($N=2$): Illustration of the two possible shapes for the error upper bound function depending on the magnitude of the singular value $\sigma_R$. Typical values of parameters $a$, $b$ $c$ were chosen as in the digital image example of Fig. \ref{Fig_Illustration_Images} ($I_1=I_2=512$ and $R=256$)}.}
   \label{FigShapeBound}
\end{figure}

It is important to note that the obtained theoretical bound is not tight in general because it is based on matrix inequalities that usually are not tight, such as the triangular inequality, i.e. $\|\A + \B\| \le \|\A\| + \|\B\|$ and other used matrix inequalities. However, it still is useful since: 
\begin{enumerate}
\item The case $\epsilon=0$ (exact case) is not realistic since always real world datasets are not low multilinear rank, so a theoretical bound, even if it is not tight is needed to characterize the reconstruction behavior in real applications; 
\item The bound, as a function of $\epsilon$ is asymptotically linear, i.e. of order $\mathcal{O}(\epsilon)$. So as more precise the model is (smaller $\epsilon$), much better reconstructions, and of the same order, are expected, which is intuitive but can be only proved by a theoretical bound. In particular, it proves that, the bound approaches zero as the best low-rank approximation error $\epsilon \to 0$, when $\tau=0$ or $\tau\propto \epsilon$, and perfect reconstruction is obtained when the signal has rank-$R$, as it was already proved in Theorem \ref{theoremExact}; 
\end{enumerate}

\subsection{Optimal selection of threshold parameter $\tau$}

A quite important question arises here: What is the optimal value of the threshold parameter $\tau$? Since the objective is to obtain a reconstruction error as small as possible, we can try to minimize the upper bound and hope that the theoretical bound gets its minimum at a point close to the minimum actual error. After a careful look at eqn. (\ref{errorbound}) as a function of $\tau$, if we define 

\begin{equation}\label{tau0}
\tau_0=\epsilon\sqrt{\frac{c}{a}}, 
\end{equation}
we can identify two different cases:
\begin{itemize}
\item {\bf CASE I (small $\sigma_R$):} When $\sigma_R <\frac{1}{2}\tau_0$, the error bound is a convex function attaining its global minimum at $\tau=\tau_0$. This means that we should use $\tau_{opt}=\tau_0$. Note that the error bound for the original reconstruction formula (\ref{reconstexact}) corresponds to the case of setting $\tau=0$, which gives us a larger error bound (see Fig. \ref{FigShapeBound} (top)).
\item {\bf CASE II (large $\sigma_R$):} On the other hand, when $\sigma_R >\frac{1}{2}\tau_0$, the best choice is to set $\tau<\sigma_R$, which corresponds to using the original reconstruction formula of eqn. (\ref{reconstexact}), i.e., with $\tau_{opt}=0$ (see Fig. \ref{FigShapeBound} (bottom)).
\end{itemize}

However, it is important to note that, in practice, we are not able to compute the optimal parameter $\tau_0$ because we do not know matrices $\A_1$ and $\A_2$ which depends on the SVD decomposition of the original unknown signal. However, we may use a rough overestimated approximation of this parameter by assuming that $\|\A_n\Ph_n\|+1\approx \|\A_n\Ph_n\|$ and using the fact that $ \|\A_n\Ph_n\| \le  \|\A_n\|\|\Ph_n\|$ which provides us the following rough estimation:
\begin{equation}
\tau_0 \lessapprox \epsilon \|\Ph_1\|\|\Ph_2\|.
\end{equation}
But, the problem still remains challenging since usually we do not know exactly the error of the best low-rank estimation $\epsilon$. 



\subsection{Error Upper Bound for a particular 3D case (N=3)}
In this section we consider the recovery of a $3$-rd order tensor $\tX\in{\R^{I_1\times I_2\times I_3}}$, for the particular and very important case where the same sensing operator is applied to every frontal slice of a tensor as considered hyperspectral CS imaging or video CS \cite{Duarte:cv}. In other words, we assume that matrix $\Ph_3 = \I\in{\R^{I_3\times I_3}}$ (i.e. $R_3=I_3$) is the identity matrix, thus the following multi-way projections are available: $\Z_1 = \X_{(1)}(\I \otimes \Ph_2)^T \in{\R^{I_1 \times R_2I_3}}$, $\Z_2 = \X_{(2)}(\I \otimes \Ph_1)^T \in{\R^{I_2 \times R_1I_3}}$ and $\Z_3 = \X_{(3)}(\Ph_2 \otimes \Ph_1)^T \in{\R^{I_3 \times R_1R_2}}$. In this case, the core tensor $\tW$ defined in eqn. (\ref{core}) becomes $\tW=\tZ^{(3)}$.
The following theorem is the $3$-rd order counterpart of Theorem \ref{theoremApprox2D} and provides an upper bound for the reconstruction error. In order to make the analysis simpler, we consider that the smallest singular values of tensor $\tW$ in mode-1 and mode-2 are the same and it is denoted as $\sigma_R = \sigma_{R_1} = \sigma_{R_2}$.

\begin{theorem} \label{theoremApprox3D}.
Let tensor $\tX\in{\R^{I_1\times I_2 \times I_3}}$ be approximated by a multilinear-rank-$(R_1,R_2,R_3)$ tensor $\tX_0\in{\R^{I_1\times I_2 \times I_3}}$, i.e. $\tX = \tX_0 + \tE$ where $\|\tE\|_F \le \epsilon$ and, given sensing matrices $\Ph_1 \in{\R^{R_1 \times I_1}}$ and $\Ph_2\in{\R^{R_2 \times I_2}}$ such that $\tW = \tX\times_1 \Ph_1 \times_2 \Ph_2$ is full-rank, then the following error upper bound holds true:
\begin{equation}\label{errorbound3D}
\|\tX - \hat{\tX}_\tau\|_F   \le  \left\{
  \begin{array}{l l}
    b \epsilon + c \frac{\epsilon^2}{\sigma_R}, & \quad \text{if $\tau \le \underline{\sigma}$}\\
   a \tau + b \epsilon + c \frac{\epsilon^2}{\tau}, & \quad \text{if $\tau > \overline{\sigma}$}
  \end{array} \right.   
\end{equation}
where $\sigma_R = \sigma_{R_1}= \sigma_{R_2}$, $\overline{\sigma} =\max{(\sigma_{R},\sigma_{R_3})}$ and $\underline{\sigma} =\min{(\sigma_{R},\sigma_{R_3})}$. (i.e., the maximum and minimum of the smallest singular values of the mode-$n$ unfolding matrices $\W_{(n)}$), and constants $a$, $b$ and $c$ are defined below.

\begin{proof}
Following the same line of reasoning as used in the proof of Theorem \ref{theoremApprox2D}, let $\tX_0 = \tGamma \times_1 \U_1 \times_2\U_2\times_3\U_3$ be the truncated HOSVD (i.e., having orthogonal factors $\U_n \in{\R^{I_n\times R_n}}$), which always exists since tensor $\tX_0$ has multilinear-rank-$(R_1,R_2,R_3)$, by defining new factors $\A_n \in{\R^{I_n\times R_n}}$ according to eqn. (\ref{basischange}) and noting that $\A_3 = \I$, we obtain:
\begin{equation}\label{result13D}
\tX_0 = \tG \times_1 \A_1 \times_2 \A_2,  \mbox{   with  } \tG = \tW - \tE \times_1 \Ph_1 \times_2 \Ph_2,
\end{equation}
where $\tW = \tX \times_1 \Ph_1 \times_2 \Ph_2$. 

The mode-$1$ measurement matrix is $\Z_1 = \X_{(1)}(\I \otimes \Ph_2)^T = (\tX_0)_{(1)}(\I \otimes \Ph_2)^T + \E_{(1)}(\I \otimes \Ph_2)^T = \A_1\W_{(1)} + \F_1$, with 
\begin{equation}\label{F1}
\F_1=(\I - \A_1\Ph_1)\E_{(1)}(\I \otimes \Ph_2)^T,
\end{equation}
and, using a similar analysis as above, we obtain that $\Z_2 =  \A_2\W_{(2)} + \F_2$, with 
\begin{equation}\label{F2}
\F_2=(\I - \A_2\Ph_2)\E_{(2)}(\I \otimes \Ph_1)^T,
\end{equation}
for mode-$2$, and $\Z_3 = \W_{(3)}$ for mode-$3$. Using these expressions, the reconstructed tensor becomes:
\begin{equation}\label{reconst3D}
\hat{\tX}_\tau =  \tOmega \times_1 (\A_1\W_{(1)} + \F_1) \times_2 (\A_2\W_{(2)} + \F_2) \times_3 \W_{(3)},
\end{equation}
where $\tOmega = \tW \times_1 \W_{(1)}^{\ast_\tau} \times_2 \W_{(2)}^{\ast_\tau} \times_3 \W_{(3)}^{\ast_\tau}$. The latter equation can be written in the following way
\begin{equation}\label{Xtensordecomp}
\hat{\tX}_\tau = \tB_1 + \tB_2 + \tB_3 + \tB_4,
\end{equation}
with 
\begin{eqnarray}
\tB_1 &=& \tOmega \times_1 \A_1\W_{(1)} \times_2 \A_2\W_{(2)}\times_3 \W_{(3)}, \label{B1} \\
\tB_2 &=& \tOmega \times_1 \A_1\W_{(1)}  \times_2  \F_2\times_3 \W_{(3)}, \label{B2} \\
\tB_3 &=& \tOmega \times_1 \F_1 \times_2 \A_2\W_{(2)}\times_3 \W_{(3)},  \label{B3}\\
\tB_4 &=& \tOmega \times_1 \F_1 \times_2  \F_2\times_3 \W_{(3)}. \label{B4}
\end{eqnarray}

By applying Lemma \ref{propTensor}, we have that $\tOmega \times_1 \W_{(1)} \times_2 \W_{(2)} \times_3 \W_{(3)} = \tW + \tH$ and, taking into account eqn. (\ref{result13D}) and the fact that $\tX = \tX_0 + \tE$, we obtain that the first term in the right-hand side in eqn. (\ref{Xtensordecomp}) can be written as follows:
\begin{equation}
\tB_1 = \tX - \tE + \tE \times_1 \A_1\Ph_1 \times_2 \A_2\Ph_2 + \tH \times_1 \A_1 \times_2 \A_2.
\end{equation}
Now, by renaming 
\begin{equation} \label{B5}
\tB_5 = - \tE + \tE \times_1 \A_1\Ph_1 \times_2 \A_2\Ph_2 + \tH \times_1 \A_1 \times_2 \A_2, 
\end{equation}
we have that $\tB_1 = \tX + \tB_5$, which implies
\begin{equation}
\hat{\tX}_\tau - \tX = \tB_2 + \tB_3 + \tB_4 + \tB_5,
\end{equation}
thus, we can find an upper bound of the approximation error by bounding from above each of the terms $\tB_2$,$\tB_3$, $\tB_4$ and $\tB_5$ in the last equation. Using the bounds developed in Appendix B, we finally arrive at eqn. (\ref{errorbound3D}), where the constants $a$, $b$ and $c$ are defined as follows:
\begin{align} 
a =& (\sqrt{R_1}+\sqrt{R_2}+\sqrt{I_3})  \|\A_1\|\|\A_2\|, \nonumber \\
b =&  1 + \|\A_1\Ph_1\|  \|\A_2\Ph_2\| + \|\A_1\|(1+\|\A_2\Ph_2 \|)\|\Ph_1\|   +  \|\A_2\|(1+\|\A_1\Ph_1 \|)\|\Ph_2\|,   \nonumber \\
c =& (1+\|\A_1\Ph_1\|)(1+\|\A_2\Ph_2\|)\|\Ph_1\| \|\Ph_2\|. \nonumber 
\end{align}
\end{proof}
\end{theorem}

\subsection{Reconstruction Sensitivity to threshold parameter $\tau$}
Theorem \ref{theoremApprox3D} shows that the Frobenius error bound of the approximation of a 3D tensor has exactly the same flavor as the bound of the spectral norm for the reconstruction of matrices (Theorem \ref{theoremApprox2D}). However, when we apply the method to natural images using random sensing matrices (see section \ref{sec:Simulations}), the reconstructions of 2D datasets are always more unstable and more sensitive to the choice of threshold parameter $\tau$, compared to the case of 3D datasets. This distinctive behavior is because, in the 2D case, the truncated MP pseudo inverse is applied to a matrix that is usually much more ill-conditioned than the ones considered in the 3D case. To be more specific, in the 2D case, $\W$ is a square random ($R\times R$) matrix, thus it tends to be ill-conditioned compared to the rectangular random matrices used in the 3D case: $\W_{(1)}\in{\R^{R_1\times R_2I_3}}$, $\W_{(2)}\in{\R^{R_2\times I_3R_1}}$ and $\W_{(3)}\in{\R^{I_3\times R_1R_2}}$.

As a consequence and, in agreement with the observed behaviors in the experiments of section \ref{sec:Simulations}, in the 2D case it is very important to use the proper threshold parameter $\tau_0\neq 0$ defined in eqn. (\ref{tau0}) in order to provide the smallest reconstruction error and low variability of the results. On the other hand, in the 3D case, the method is always very stable with small reconstruction errors obtained with $\tau=0$, i.e. computing the standard MP pseudo inverse without truncation. This is a clear advantage of considering a number of dimensions $N > 2$ because, usually, the truncation parameter $\tau$ does not need to be tuned and optimal results are obtained by using just $\tau=0$.

\section{Numerical Simulations of CS Using Real-world Datasets}\label{sec:Simulations}
Here, we analyze various aspects of the proposed CS reconstruction method through numerical simulations and compare the performance with 2D and 3D datasets. All the simulations were performed using Matlab software on an iMac desktop computer, equipped with an Intel Core i5 processor (2.2 GHz) and 8 GB RAM. Matlab codes to reproduce the simulation results are available at \url{http://web.fi.uba.ar/~ccaiafa/Cesar/Low-Rank-Tensor-CS.html}. In order to evaluate the quality of the reconstructions we use the Peak Signal to Noise Ratio defined as follows: PSNR (dB)$=20\log_{10}{( \max(\tX)/\|\hat{\tX} - \tX\|_F}$).

\subsection{Selected illustrative examples for 2D and 3D signals}
\begin{figure}[!t]
 \centering
 \centerline{\includegraphics[width=8.0cm]{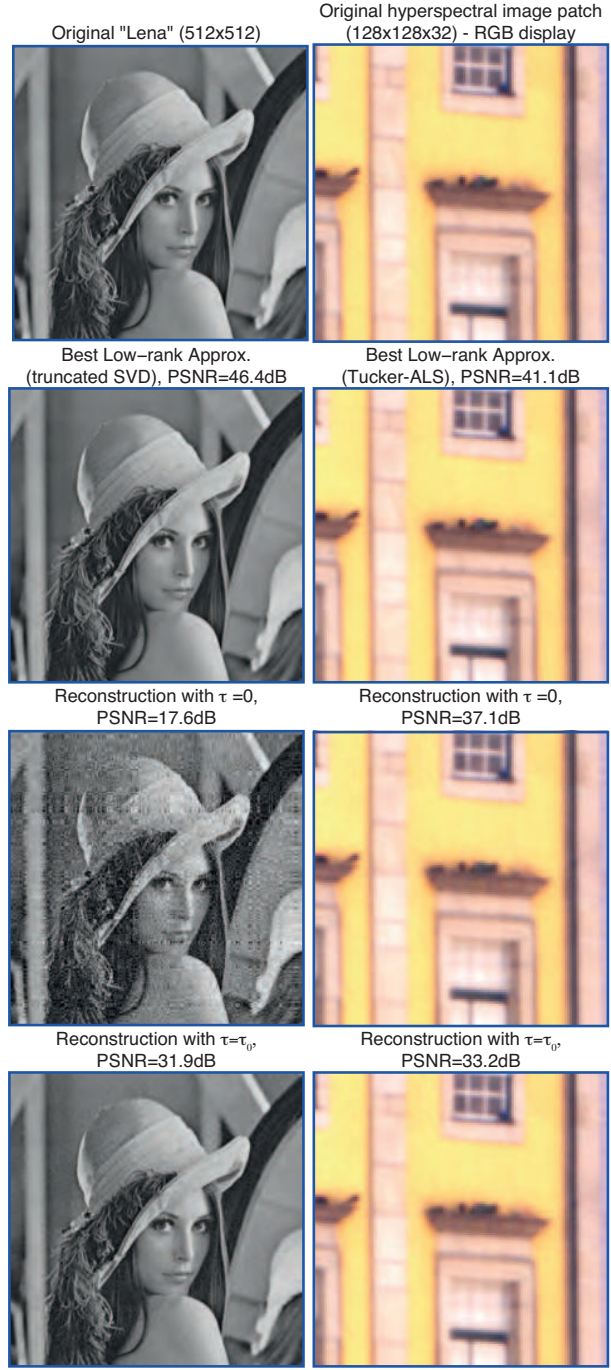}}
 \caption{\footnotesize{Original 2D and 3D datasets (top), their best low-rank approximations (2nd row) and the reconstructions obtained by our method using $\tau=0$ (3rd row) and $\tau=\tau_0$ (bottom). It is highlighted that, unlike in the 2D case, we can use $\tau_{opt}=0$ as the optimal value (lower error bound) in the 3D case.}}
   \label{Fig_Illustration_Images}
\end{figure}

First, we consider the 2D digital image ``Lena'' encoded in matrix $\X\in{\R^{512\times 512}}$ and assume that compressive Gaussian samples are collected for rows and columns as follows: $\Z_1 = \X\Ph_2^T$ and $\Z_2^T = \Ph_1\X$ with $\Ph_{1},\Ph_{2}\in{\R^{256 \times 512}}$ being matrices with independent Gaussian entries. Note that, in this case, we have $I_1=I_2=512$, $R=256$ and the core matrix $\W=\Ph_1 \X \Ph_2^T$ can be computed from the available measurement matrix $\Z_1$ or $\Z_2$ multiplying it by the appropriate sensing matrix $\Ph_1$ or $\Ph_2$, respectively. In Fig. \ref{Fig_Illustration_Images}, the original digital image is shown in the top-left panel, and right down below, its best rank-$R$ approximation (truncated SVD) and the reconstructions obtained with $\tau=0$ and $\tau=\tau_0$, as defined in eqn. (\ref{tau0}), are also shown. It is clear that the optimal result is obtained with $\tau_{opt}=\tau_0=0.19$ with PSNR=$31.9$dB, which is considerably much higher than the value obtained with $\tau=0=0.19$ (PSNR=$17.6$dB).

We also considered a 3D hyperspectral tensor image $\tX\in{\R^{128\times 128\times 32}}$, which is actually a patch extracted from ``Ribeira'', a large hyperspectral image included in a public dataset used in \cite{FOSTER:2004vi}. We also assumed Gaussian sensing matrices $\Ph_{1},\Ph_{2}\in{\R^{64 \times 128}}$ which provides us with the following measurement matrices: $\Z_1 = \X_{(1)}(\I \otimes \Ph_2)^T$, $\Z_2 = \X_{(2)}(\I \otimes \Ph_1)^T$ and $\Z_3 = \X_{(3)}(\Ph_2 \otimes \Ph_1)^T$. Note that, in this case, $\W_{(3)} = \Z_{(3)}$. In Fig. \ref{Fig_Illustration_Images}, the original tensor is shown in the top-right panel as a color RGB display, and right down below, its best multilinear-rank-$(64,64,32)$ approximation is shown, which was obtained by applying the Tucker Alternating Least Squares (ALS) algorithm of the Tensor Toolbox \cite{tensortoolbox}. We also show that the obtained reconstructions using $\tau=0$ (PSNR=$37.1$dB) is slightly better than the one obtained with $\tau=\tau_0=0.2$ (PSNR=$33.2$dB). 

As it is illustrated in these examples, in general, we can say that, working with higher dimensions is an advantage because involved matrices tend to be better conditioned avoiding the difficulty of estimating the optimal $\tau$ and allowing us to use just $\tau_{opt}=0$ (regular MP pseudo inverse). 

\subsection{Sensitivity to parameter $\tau$ analysis in the 2D and 3D cases}
In order to analyze the behavior of the reconstructions as a function of the threshold parameter $\tau$ in the 2D and 3D cases, we have generated a matrix signal by using the best rank-$R$ approximation of ``Lena'' image ($512\times 512$) and a tensor signal by using the best multilinear-rank-($R,R,I_3$) approximation (Tucker-ALS) of a tensor patch used in the previous section. By adding some Gaussian noise to these base signals, we obtained our models: $\X = \X_0 + \epsilon\E$  (2D case) and $\tX = \tX_0 + \epsilon\tE$ (3D case), where $\|\E\| = \|\tE\|_F=1$ and $\epsilon$ can be controlled by adjusting the variance of the added noise. We have normalized the matrix $\X$ and tensor $\tX$ in order to have $\|\X\|=\|\tX\|_F=1$. 

\begin{figure}[!t]
 \centering
 \centerline{\includegraphics[width=12cm]{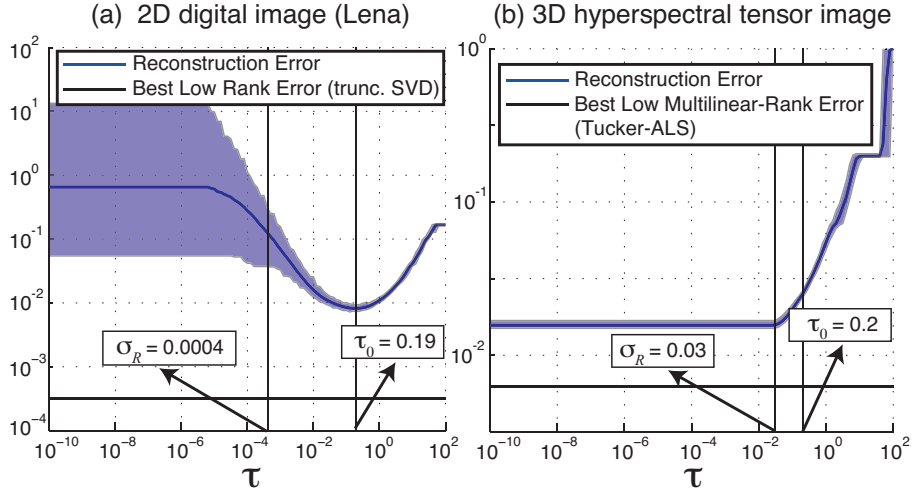}}
 \caption{\footnotesize{Approximation errors computed on 500 simulations for ``Lena'' ($512\times 512$) (a) and a ($128\times 128\times 32$)-patch of ``Ribeira''. The range of obtained values are shown as shaded areas. It is highlighted that the standard deviation of the error, at the minimum, is $3.0e-4$ in both cases ((a) and (b)), which means that the method is very robust to the actual choice of random sensing matrices $\Ph_1$ and $\Ph_2$. Averaged $\sigma_R$ and $\tau$ values are also indicated.}}
   \label{Fig_Vs_Tau}
\end{figure}

In Fig. \ref{Fig_Vs_Tau} (a), the reconstruction errors $\|\X-\hat{\X}_\tau\|$, obtained for a fixed $\epsilon=3.1\times 10^{-4}$, over a total number of $500$ simulations (with different sensing matrices in each simulation) are shown for the 2D case and compared against the best low-rank approximation (truncated SVD). We can see that the actual error has a convex shape attaining its minimum approximately at $\tau_{opt} = \tau_0=0.19$ (optimal choice). In this case, we obtained an average $\sigma_R=4.02\times 10^{-4}$ which means that matrix $\W$ is usually ill-conditioned and $\sigma_R\ll 0.5\tau_0$.

In Fig. \ref{Fig_Vs_Tau} (b) the reconstruction errors for the 3D case, $\|\tX-\hat{\tX}_\tau\|$, for a fixed $\epsilon=6.2\times 10^{-3}$ and the best low multilinear-rank, are shown. In the latter, we observe a totally different behavior compared to the 2D case, instead of a convex function, now the minimum error is obtained by choosing $\tau_{opt} = 0$, which is also slightly lower than using $\tau= \tau_0=0.2$. It is noted that, in this case, the average smallest singular value was $\sigma_R=0.03$, which means that matrices $\W_{(1)}$ and $\W_{(2)}$ are better conditioned than in the 2D case. 

Another remarkable property of the proposed reconstruction method is that it provides very low variance of the results (robustness). In the 2D case, the standard deviation s.d. is minimal at $\tau_{opt}=\tau_0$ ($s.d. = 3.0\times 10^{-4}$) and increases significantly for $\tau < \tau_0$. On the other side, the standard deviation of the errors obtained in the 3D case is very small ($s.d.= 3.0\times 10^{-4}$) and approximately constant for all the range of the parameter $\tau$ values. This means, that our method is robust to the actual random sensing matrices $\Ph_1$ and $\Ph_2$ and, the results in the 2D case is very sensitive to the choice of parameter $\tau$ opposed to the 3D case.

\subsection{Reconstruction error versus model error $\epsilon$}
\begin{figure}[!t]
 \centering
 \centerline{\includegraphics[width=12cm]{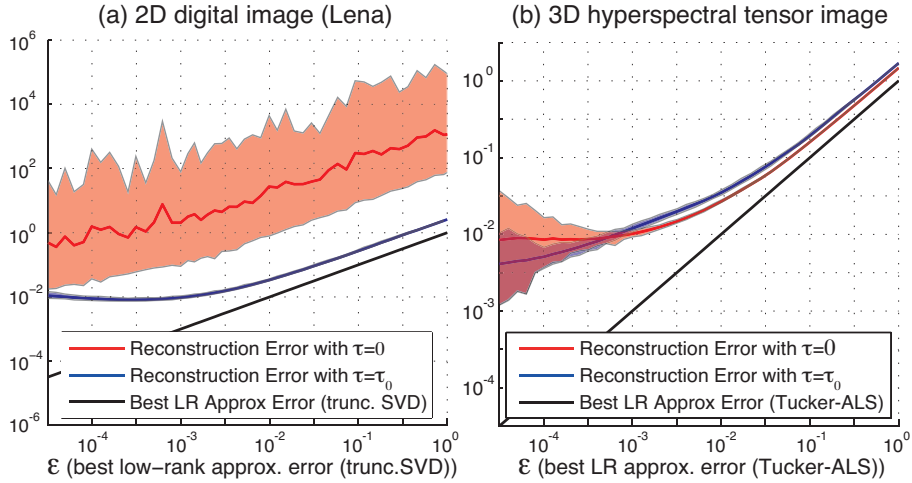}}
 \caption{\footnotesize{Approximation errors versus $\epsilon$ for 2D case (a) and 3D case (b). It is highlighted that, while $\tau=0$ gives poorer results (large errors and large deviations) compared to $\tau=\tau_0$ for the 2D case, in the 3D case, similar good results (small errors and deviations) are obtained for both, $\tau=0$ and $\tau=\tau_0$.}}
   \label{Fig_Vs_Eps}
\end{figure}

In Fig. \ref{Fig_Vs_Eps}, using the same model as in the previous section for generating 2D and 3D signals, the performance of the reconstructions using $\tau=\tau_0$ is compared against the case of using $\tau=0$, as a function of $\epsilon$ (best low-rank approximation error). In the 2D case (a), the original reconstruction formula (eqn. (\ref{reconstexact})) gives always poorer results, i.e., larger errors, and less robust behavior (larger deviations over repeated simulations). On the other side, when using the modified formula (eqn. (\ref{modified_reconst})) with the optimal threshold parameter $\tau_{opt}=\tau_0=\epsilon\sqrt{c/a}$, the reconstruction errors are robust and much smaller, in fact they are  close to the case of the best low-rank approximation given by the truncated SVD for $\epsilon > 0.003$, approximatelly. 

On the other hand, in the 3D case (b), robust (small deviations over repeated simulations) and small errors are obtained with $\tau=0$, as well as with $\tau=\tau_0$. Thus, our method is less sensitive to the choice of parameter $\tau$ in the 3D case compared to the 2D case. Also, for a very small model noise ($\epsilon < 3.1\times 10^{-4}$), better reconstructions are obtained with $\tau=\tau_0$ while, for a larger model noise ($\epsilon > 3.1\times 10^{-4}$), using $\tau=0$ provides slightly better reconstructions.

\begin{figure}
 \centering
 \centerline{\includegraphics[width=11cm]{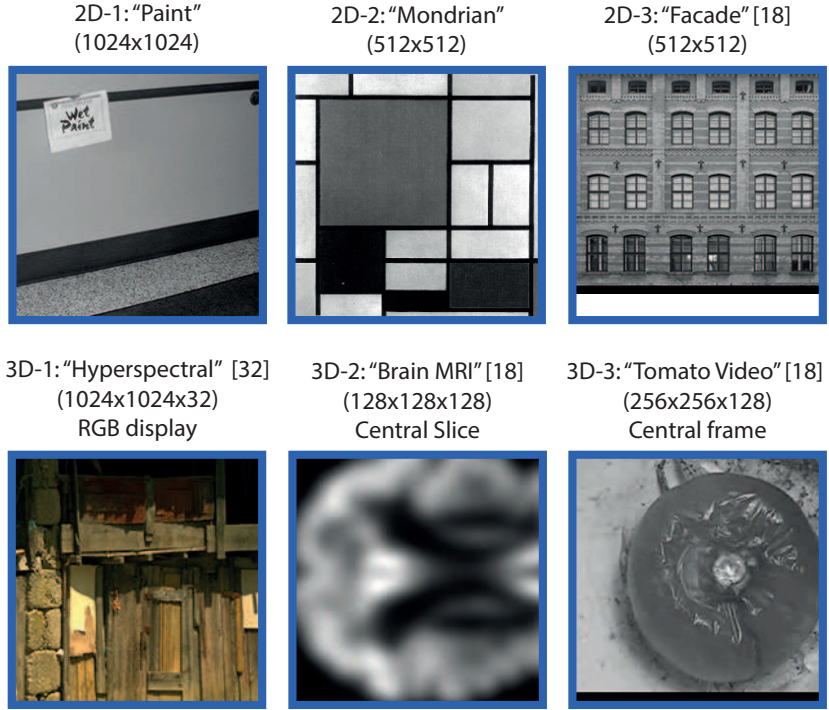}}
 \caption{\footnotesize{2D - images (top) and 3D tensors (bottom) used in our simulations. 3D-1 hyperspectral image corresponds to the scene ``'Farme' included in a public dataset (\url{http://personalpages.manchester.ac.uk/staff/david.foster/})\cite{FOSTER:2004vi}. 3D-2 Brain MRI and 3D-3 Tomato Video datasets were used in the paper \cite{JiLiu:bh} and can be obtained from the author's webpage.}}
   \label{Fig_datasets}
\end{figure}

\begin{figure}[!t]
 \centering
 \centerline{\includegraphics[width=13cm]{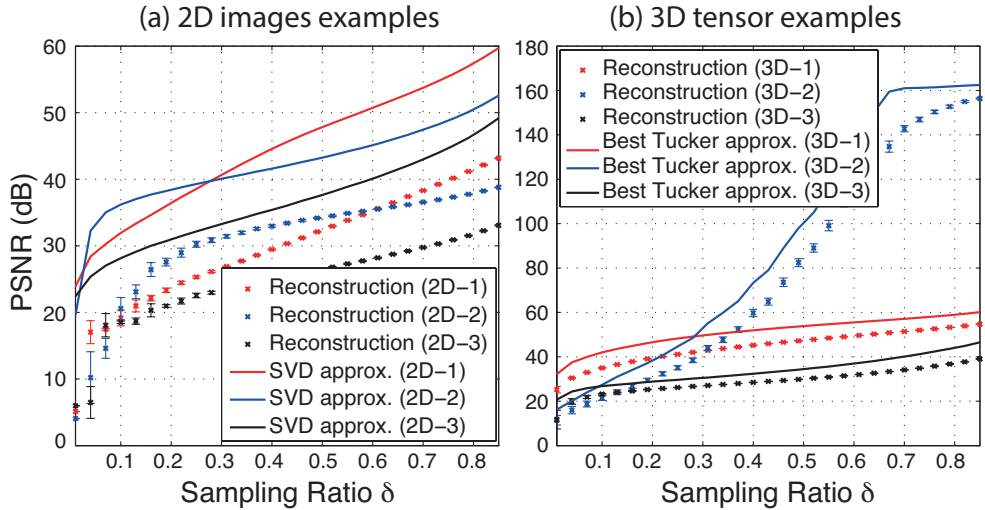}}
 \caption{\footnotesize{Performance of reconstructions for 2D (a) and 3D (b) datasets. Mean values plus/minus the standard deviations over $100$ Monte Carlo simulations are shown. PSNRs associated with the best low rank (truncated SVD) and the best multilinear-rank (obtained through the Tucker-ALS algorithm) approximations are also shown for reference.}}
   \label{Fig_Vs_SR}
\end{figure}

\subsection{Reconstruction error versus sampling ratio $\delta$}
In Fig. \ref{Fig_Vs_SR}, we analyze, through $100$ Monte Carlo simulations, the approximation errors obtained by our method and we compare it to the best possible low-rank approximation as a function of the sampling ratio $\delta$, i.e. the size of the non-redundant measurement data divided by the size of the original dataset, for several 2D ($I\times I$) images  (see Fig. \ref{Fig_datasets}-top) and different types of 3D ($I\times I \times I_3$) data tensors: a hyperspectral image \cite{FOSTER:2004vi}, a Magnetic Resonance Image (MRI) of a brain \cite{JiLiu:bh}, and a video squence \cite{JiLiu:bh} (see Fig. \ref{Fig_datasets}-bottom). 

We note that, to compute the reconstruction, matrices $\Z_1$, $\Z_2$ and $\W$ are redundant, in fact matrix $\W$ can be computed directly from measurements $\Z_1$ as follows: $\W=\Ph_1 \Z_1$. Also, by assuming the following block notation: $\Z_1^T = (\Z_{1,1}, \Z_{1,2})$, $\Z_2^T = (\Z_{2,1}, \Z_{2,2})$, $\Ph_1 = (\Ph_{1,1}, \Ph_{1,2})$ and $\Ph_2 = (\Ph_{2,1}, \Ph_{2,2})$, with $\Z_{1,1}, \Z_{2,1}, \Ph_{1,1}, \Ph_{2,1}  \in{\R^{R\times (I-R)}}$ and $\Z_{1,2}, \Z_{2,2}, \Ph_{1,2}, \Ph_{2,2}  \in{\R^{R\times R}}$, it is not difficult to prove that, in the 2D case, $\Z_{1,2}$ can be actually computed from $\Z_2$ and $\Z_{1,1}$ as follows:
\begin{equation}
\Z_{1,2} = (\Ph_2\Z_2 - \Z_{1,1}\Ph_{1,1}^T)(\Ph_{1,2}^{-1})^{T},
\end{equation}
where matrix $\Ph_{1,2}$ is assumed to be invertible. Thus, the size of the minimal non-redundant measurement data required is given by $2RI - R^2$ (i.e. the sum of the sizes of matrices $\Z_2$ and $\Z_{1,1}$), which means that the sampling ratio can be defined as follows:
\begin{equation}\label{SR}
\delta = 2\frac{R}{I} - \left(\frac{R}{I}\right)^2.
\end{equation}
It is easy to see that, the same sampling ratio formula is also valid for the 3D case when the sensing matrix in the mode-$3$ is equal to the identity matrix.

In Fig. \ref{Fig_Vs_SR} (a), the obtained PSNRs using $\tau=\tau_0$ for all the 2D datasets are shown while, in Fig. \ref{Fig_Vs_SR} (b), the obtained PSNRs using $\tau=0$ for all the 3D datasets are shown. These results demonstrate the robustness of the method regarding the actual selection of sensing matrices $\Ph_1,\Ph_2$, since the observed variance of PSNR is very small. It is noted that, the gap between the actual PSNR and the best (ideal) Low-Rank (LR) approximation is much smaller in the 3D case compared to the 2D case. In particular, dataset 3D-2 (Brain MRI), showed a very strong low rank structure allowing us to obtain very good reconstructions compared to the other two datasets. For instance, for $\delta=0.4$, the obtained PSNR is around $60$dB for dataset 3D-2, $45$dB for dataset 3D-1 (hyperspectral image) and $28$dB for dataset 3D-3 (video sequence).


\subsection{Comparison against state-of-the-arts sparsity based algorithms} \label{sec:exp_comparison}
Here we compare our proposed reconstruction method against state-of-the-arts sparsity based CS algorithms for multidimensional datasets using random sensing matrices of types: a) Gaussian, and b) Bernoulli, i.e. with entries being $+1$ or $-1$ with equal probability. 

In Table \ref{table_Ex2D} (rightmost columns), we show the obtained PSNRs by applying our method to all the 2D datasets shown in Fig. \ref{Fig_datasets} (top) using $R=0.2I$, and we compare them against the results obtained by applying the Kronecker-CS (leftmost columns), consisting in using the SPGL1 algorithm \cite{Duarte:cv,Rivenson:2009p285} and taking into account the Kronecker structure using Daubechies Wavelet bases. At first glance, it would seem that our proposed method uses more measurements than Kronecker-CS algorithms because we use two tensor measurements, i.e. $\underline{\mathbf{Z}}^{(1)}$, $\underline{\mathbf{Z}}^{(2)}$, instead of only one tensor $\underline{\mathbf{W}}\in{\R^{R\times R}}$. In order to make a more complete and fair comparison, in the central columns of Table \ref{table_Ex2D}, we show the results of applying the Kronecker-CS method using a larger $R$ in order to attain the same sampling ratio $\delta = 0.36$ as in our method. It is interesting to note that, even in the case of having the same sampling ratio, our method gives the best results in all cases except with the 2D-1 ``Paint'' dataset, indicating that, its sparse model is richer than its low-rank model. On the other hand, datasets 2D-2 and 2D-3 have stronger low rank structures, which allows our proposed method to obtain better reconstructions.

In Table \ref{table_Ex3D}, we show the obtained PSNRs by applying our method to the 3D datasets shown in Fig. \ref{Fig_datasets} (bottom), and we compare them against the results obtained by applying the N-Way Block Sparse Orthogonal Matching Pursuit (NBOMP) algorithm developed in \cite{Caiafa:2012iv} for the same $R=0.125$ (leftmost columns) and same sampling ratio $\delta=0.23$ (central columns). Note that, for example, the hyperspectral image in dataset 3D-1 is so large that it is almost impossible to apply other CS algorithms, even Kronecker-CS methods using a standard computer. 

In Table \ref{table_TimeEx2D} and Table \ref{table_TimeEx3D}, we compare the computation times required in each case showing that our direct method provides an extremely faster computation. For a fixed sampling ratio $\delta$, our method is 5 orders of magnitude faster in the 2D case, and 2 orders of magnitude faster in the 3D case. For example, dataset 2D-1 ``Paint'' required almost one hour to be reconstructed by using the Kronecker-CS algorithm and our method takes only 41 milliseconds to build the reconstruction.

\begin{table}
\begin{minipage}[!t]{1.0\textwidth}
\centering
\caption{\footnotesize{Reconstruction quality (PSNR) of the proposed method and the Kronecker-CS algorithm \cite{Duarte:cv} for 2D signals using Gaussian and Bernoulli sensing matrices.}}
\scalebox{0.95}{\begin{tabular}{ | c | c | c | c | c | c | c | }
\hline
{}							& \multicolumn{2}{c|}{\textbf{Kronecker-CS}} 	&  \multicolumn{2}{c|}{\textbf{Kronecker-CS}}  		&  \multicolumn{2}{c|}{\textbf{New Method}}  \\ 
{}							& \multicolumn{2}{c|}{$R=0.2I$} 	&  \multicolumn{2}{c|}{$R=0.6I$}  	&  \multicolumn{2}{c|}{$R=0.2I$}  \\ 
{} 							& \multicolumn{2}{c|}{$\delta = 0.04$} 	&  \multicolumn{2}{c|}{$\delta = 0.36$}  		&  \multicolumn{2}{c|}{$\delta = 0.36$}  \\ 
\hline
\textbf{Data}						 & \textbf{Gaussian} 		& \textbf{Bernoulli} 		& \textbf{Gaussian}  	& \textbf{Bernoulli} 		& \textbf{Gaussian}  	& \textbf{Bernoulli} \\
\hline
\textbf{2D-1}	 					& $22.8$dB			& $22.3$dB			& ${\bf31.7}$dB			& $30.6$dB		& $28.3$dB			& $28.4$dB \\
\hline
\textbf{2D-2} 						& $9.8$dB				& $12.6$dB			& $26.8$dB			& $26.6$dB		& ${\bf32.5}$dB			& $32.3$dB \\
\hline
\textbf{2D-3}						& $13.9$dB			& $12.7$dB			& $21.9$dB			& $21.7$dB		& $24.0$dB			& ${\bf24.2}$dB \\
\hline
\end{tabular}
}
\label{table_Ex2D}
\end{minipage}
\end{table}

\begin{table}
\begin{minipage}[!t]{1.0\textwidth}
\centering
\caption{\footnotesize{Reconstruction quality (PSNR) of the proposed method and N-way Block-Sparse OMP (NBOMP) algorithm \cite{Duarte:cv} for 3D signals using Gaussian and Bernoulli sensing matrices.}}
\scalebox{0.95}{\begin{tabular}{ | c | c | c | c | c | c | c | }
\hline
{}							& \multicolumn{2}{c|}{\textbf{NBOMP-CS}} 	&  \multicolumn{2}{c|}{\textbf{NBOMP-CS}}  		&  \multicolumn{2}{c|}{\textbf{New Method}}  \\ 
{}							& \multicolumn{2}{c|}{$R=0.125I$} 	&  \multicolumn{2}{c|}{$R=0.48I$}  	&  \multicolumn{2}{c|}{$R=0.125I$}  \\ 
{} 							& \multicolumn{2}{c|}{$\delta = 0.016$} 	&  \multicolumn{2}{c|}{$\delta = 0.23$}  		&  \multicolumn{2}{c|}{$\delta = 0.23$}  \\ 
\hline
\textbf{Data}						 & \textbf{Gaussian} 		& \textbf{Bernoulli} 		& \textbf{Gaussian}  	& \textbf{Bernoulli} 		& \textbf{Gaussian}  	& \textbf{Bernoulli} \\
\hline
\textbf{3D-1}	 					& $21.9$dB			&  $22.4$dB			& $39.8$dB			&  $39.9$dB		& ${\bf40.6}$dB		&  $40.5$dB \\
\hline
\textbf{3D-2}	 					& $6.3$dB				&  $6.1$dB			& $27.7$dB			&  $29.2$dB		& $33.0$dB		&  ${\bf34.4}$dB \\
\hline
\textbf{3D-3}	 					& $3.6$dB				&  $3.6$dB			& $23.1$dB			&  $22.2$dB		& $25.8$dB		&  ${\bf25.7}$dB \\
\hline
\end{tabular}
}
\label{table_Ex3D}
\end{minipage}
\end{table}

\begin{table}
\begin{minipage}[!t]{1.0\textwidth}
\centering
\caption{\footnotesize{Computational time comparison between the proposed method and Kronecker-CS algorithm \cite{Duarte:cv} for 2D signals.}}
\scalebox{0.95}{\begin{tabular}{ | c | c | c | c | c | c | c | }
\hline
{}							& \multicolumn{1}{c|}{\textbf{Kronecker-CS}} 	&  \multicolumn{1}{c|}{\textbf{Kronecker-CS}}  		&  \multicolumn{1}{c|}{\textbf{New Method}}  \\ 
{}							& \multicolumn{1}{c|}{$R=0.2I$} 			&  \multicolumn{1}{c|}{$R=0.6I$}  				&  \multicolumn{1}{c|}{$R=0.2I$}  \\ 
{} 							& \multicolumn{1}{c|}{$\delta = 0.04$} 		&  \multicolumn{1}{c|}{$\delta = 0.36$}  			&  \multicolumn{1}{c|}{$\delta = 0.36$}  \\ 
\hline
\textbf{Data}						 		& \textbf{Time (sec.)} 		 	& \textbf{Time (sec.)} 		  	& \textbf{Time (sec.)} \\
\hline
\textbf{2D-1}	 								& $5.4\times 10^{2}$					& $3.5\times 10^3$					& $4.1\times 10^{-2}$ \\
\hline
\textbf{2D-2} 									& $1.2\times 10^{2}$					& $6.7\times 10^{2}$					& $8.8\times 10^{-3}$ \\
\hline
\textbf{2D-3}									& $1.2\times 10^{2}$					& $7.7\times 10^{2}$					& $9.3\times 10^{-3}$ \\
\hline
\end{tabular}
}
\label{table_TimeEx2D}
\end{minipage}
\end{table}

\begin{table}
\begin{minipage}[!t]{1.0\textwidth}
\centering
\caption{\footnotesize{Computational time comparison between the proposed method and N-way Block-Sparse OMP (NBOMP) algorithm \cite{Caiafa:2012iv} for 3D signals.}}
\scalebox{0.95}{\begin{tabular}{ | c | c | c | c | c | c | c | }
\hline
{}							& \multicolumn{1}{c|}{\textbf{NBOMP-CS}} 	&  \multicolumn{1}{c|}{\textbf{NBOMP-CS}}  		&  \multicolumn{1}{c|}{\textbf{New Method}}  \\ 
{}							& \multicolumn{1}{c|}{$R=0.125I$} 			&  \multicolumn{1}{c|}{$R=0.48I$}  				&  \multicolumn{1}{c|}{$R=0.125I$}  \\ 
{} 							& \multicolumn{1}{c|}{$\delta = 0.016$} 		&  \multicolumn{1}{c|}{$\delta = 0.23$}  			&  \multicolumn{1}{c|}{$\delta = 0.23$}  \\ 
\hline
\textbf{Data}						  		& \textbf{Time (sec.)} 		  	& \textbf{Time (sec.)} 		  	& \textbf{Time (sec.)} \\
\hline
\textbf{3D-1}	 								& $6.8$						& $4.0\times 10^{3}$					& $1.1\times 10^{1}$ \\
\hline
\textbf{3D-2}	 								& $8.2$						& $1.8\times 10^{1}$					& $4.4\times 10^{-1}$ \\
\hline
\textbf{3D-3}	 								& $4.1\times 10^{1}$				& $2.4\times 10^{2}$					& $1.8$ \\
\hline
\end{tabular}
}
\label{table_TimeEx3D}
\end{minipage}

\end{table}

\section{Conclusions and Discussion} \label{sec:Conclusions}
We have provided a new CS reconstruction formula for multidimensional signals assuming that a set of multi-way projections are available and that a good low multilinear-rank approximation exists. Compared to existing sparsity based CS methods, our model does not require to assume sparsity neither a dictionary based representation. 

In sparsity based CS, \cite{Rivenson:2009p285,Duarte:cv,Caiafa:2012iv}, available theoretical guarantees are based on sparsity levels and properties of the sensing/dictionary matrix such as RIP (restricted isometry property) and coherence. It is known that, if a signal has an exact sparse representation, then classical CS algorithms such as Matching Pursuit (MP) or Basis Pursuit (BP) are able to provide an exact reconstruction and, if the signal has an approximate sparse representation, then MP and BP provide reconstructions that are stable, i.e., they are close to the original signal with high probability.
In our approach, we prove that, if the signal has an exact low multilinear-rank representation, then the proposed reconstruction is exact and, in the realistic case that the signals have only approximate low multilinear-rank representations, the reconstruction error has an upper bound that is of order $\mathcal{O}(\epsilon)$ where $\epsilon$ is the model error.

Our simulation results showed that, the present method has the following significant advantages: 
\begin{enumerate}
\item It is {\it super fast} because it does not involve iterations making it potentially suitable for large-scale problems; 
\item It is {\it stable}, in the sense that tensors which are well approximated by a Tucker model, are also well reconstructed by the proposed method,
\item It is {\it robust} because the reconstruction performance is not sensitive to the used Gaussian/Bernoulli sensing matrices. 
\end{enumerate}
Moreover, we have shown that, working with higher dimensions ($N>2$), in particular with $N=3$, gives us additional advantages because the involved matrices are better conditioned compared to the 2D case, providing more stable results and allowing us to use the standard MP pseudo inverse, i.e. without truncation. 

We have shown the applicability of our method to the case of hyperspectral compressive imaging for which the technology is already available \cite{Duarte:cv,Robucci:2010cw} and applied it also to other kinds of 3D datasets such as MRI images and video sequences. 

Our model is presented in a general setting which makes it hopefully applicable to next generation of multidimensional sensors and new methods for big-data processing under the assumption of the existence of a good multilinear rank representation and the availability of multilinear compressive measurements.

\appendices

\section*{Appendix A: Proof of Lemma \ref{lemmaTensor}} \label{sec:LemmaProof}

Let us consider the mode-$1$ unfolding of the left-hand side of eqn. (\ref{lefthandLemma}), and apply the property (\ref{PropI}) to matrix $\W_{(1)}$, i.e.  $\W_{(1)}\W_{(1)}^{\ast_\tau}\W_{(1)} = \W_{(1)} + \mathbf{P}_{(1)}$, then we obtain:
\begin{gather*}
\W_{(1)}\W_{(1)}^{\ast_\tau}\W_{(1)}(\W_{(3)}\W_{(3)}^{\ast_\tau}\otimes \W_{(2)}\W_{(2)}^{\ast_\tau})^T = \\
\W_{(1)}(\W_{(3)}\W_{(3)}^{\ast_\tau}\otimes \W_{(2)}\W_{(2)}^{\ast_\tau})^T + \mathbf{P}_{(1)}(\W_{(3)}\W_{(3)}^{\ast_\tau}\otimes \W_{(2)}\W_{(2)}^{\ast_\tau})^T.
\end{gather*}
The mode-$2$ representation of the last equation is as follows:
\begin{gather*}
\W_{(2)}\W_{(2)}^{\ast_\tau}\W_{(2)}(\W_{(3)}\W_{(3)}^{\ast_\tau}\otimes \I)^T + \W_{(2)}\W_{(2)}^{\ast_\tau}\mathbf{P}_{(2)}(\W_{(3)}\W_{(3)}^{\ast_\tau}\otimes \I)^T.
\end{gather*}
Now, by applying the property (\ref{PropI}) to matrix $\W_{(2)}$, i.e. $\W_{(2)}\W_{(2)}^{\ast_\tau}\W_{(2)} = \W_{(2)} + \Q_{(2)}$, and using the mode-$3$ representation of the last equation, we arrive at:
\begin{gather*}
\W_{(3)}\W_{(3)}^{\ast_\tau}\W_{(3)} + \W_{(3)}\W_{(3)}^{\ast_\tau} \Q_{(3)} + \W_{(3)}\W_{(3)}^{\ast_\tau}\mathbf{P}_{(3)}(\W_{(2)}\W_{(2)}^{\ast_\tau}\otimes \I)^T.
\end{gather*}
Using again the property (\ref{PropI}), i.e. $\W_{(3)}\W_{(3)}\W_{(3)}^{\ast_\tau} = \W_{(3)} + \mathbf{R}_{(3)}$, and writing the last equation in tensor format, we finally obtain: $\tW + \tH$, with 
\begin{equation*}
\tH = \tR + \tQ\times_3 \W_{(3)}\W_{(3)}^{\ast_\tau} + \tP \times_2 \W_{(2)}\W_{(2)}^{\ast_\tau} \times_3 \W_{(3)}\W_{(3)}^{\ast_\tau}, 
\end{equation*}
which, by using elementary properties of the Frobenius and spectral norms and the fact that $\|\P_{(1)}\|$, $\|\Q_{(2)}\|$, $\|\mathbf{R}_{(3)}\| \le \tau$ ($\tau > \overline{\sigma}$) and $\|\P_{(1)}\|=\|\Q_{(2)}\|=\|\mathbf{R}_{(3)}\| =0$ ($\tau \le \underline{\sigma}$), implies that $\|\tH\|_F \le \sqrt{R_3}\tau + \sqrt{R_2}\tau + \sqrt{R_1}\tau$ if $\tau > \overline{\sigma}$, and $\tH = \t0$ if $\tau \le \underline{\sigma}$. 

\section*{Appendix B: bounds for tensors $\tB_2$, $\tB_3$, $\tB_4$ and $\tB_5$} \label{sec:AppBounds}
By considering the mode-$2$ unfolding of eqn. (\ref{B2}) we have that
\begin{equation*}
\|\tB_2\|_F = \|\F_2 \W_{(2)}^{\ast_\tau}\W_{(2)}(\W_{(3)}\W_{(3)}^{\ast_\tau}\otimes \A_1\W_{(1)}\W_{(1)}^{\ast_\tau})^T\|_F,
\end{equation*}
from which, by applying the property (\ref{PropIII}) of section \ref{sec:MoorePenrose}, and using the fact that $\| \A \B \| \leq \| \A \| \| \B \|_F$, that spectral norm is a sub-multiplicative norm, and that the spectral norm of the Kronecker product of $\A$ and $\B$ is equal to the product of the norm of $\A$ and that of $\B$, namely, $\|\A \otimes \B\|=\|\A\|\|\B\|$, we obtain that $\|\tB_2\|_F \le \|\A_1\| \|\F_2\|_F$. Now, using the definition of matrix $\F_2$ (see eqn. (\ref{F2})), we have that $\|\F_2\|_F \le \|\I - \A_2\Ph_2\| \|\I\otimes \Ph_1\| \| \tE\|_F$ which implies the following bound for $\tB_2$
\begin{equation}
\|\tB_2\|_F \le \|\A_1\| (1 + \|\A_2\Ph_2\|)\|\Ph_1\|\epsilon.
\end{equation}
Analogously, we obtain the following bound for $\tB_3$:
\begin{equation}
\|\tB_3\|_F \le \|\A_2\| (1 + \|\A_1\Ph_1\|)\|\Ph_2\|\epsilon.
\end{equation}
By considering the mode-$1$ representation of eqn. (\ref{B4}) we obtain the following inequality:
\begin{equation*}
\|\tB_4\|_F \le \|\F_1 \W_{(1)}^{\ast_\tau}\W_{(1)}\|_F \| \W_{(3)}\W_{(3)}^{\ast_\tau} \otimes \F_2\W_{(2)}^{\ast_\tau}\|_F,
\end{equation*}
and using properties of the MP pseudo-inverse and the already obtained bounds for $\|\F_1\|_F$ and $\|\F_2\|_F$ we finally arrive at:
\begin{equation*}
\|\tB_4\|_F \le (1 + \|\A_1\Ph_1\|) (1 + \|\A_2\Ph_2\|) \frac{\|\Ph_1\|\|\Ph_2\|\epsilon^2}{max{(\tau,\sigma_R)}}.
\end{equation*}
From the definition of $\tB_5$ in eqn. (\ref{B5}) and by using Lemma \ref{lemmaTensor}, we derive the following bounds: 
$\|\tB_5\|_F   \le \epsilon (1 + \|\A_1\Ph_1\| \|\A_2\Ph_2\|) + \tau (\sqrt{R_1}+\sqrt{R_2}+\sqrt{I_3})\|\A_1\|\|\A_2\|$, if $\tau > \overline{\sigma}$, and 
$\|\tB_5\|_F   \le \epsilon (1 + \|\A_1\Ph_1\| \|\A_2\Ph_2\|)$, if $\tau \le \underline{\sigma}$.



\ifCLASSOPTIONcompsoc
  \section*{Acknowledgments}
\else
  \section*{Acknowledgment}
\fi
The authors thank anonymous Reviewers for their insightful critical comments and useful suggestions. This work was partially supported by grants from ANPCyT (Agencia Nacional de Promoci\'{o}n Cient\'{i}fica y Tecnol\'{o}gica), PICT 2012 \#1519; and CONICET (Consejo Nacional de Investigaciones Cient\'{i}ficas y T\'{e}cnicas), PIP \#114-201101-00021.

\ifCLASSOPTIONcaptionsoff
  \newpage
\fi



\bibliographystyle{IEEEtran}
\bibliography{Caiafa}
\end{document}